\documentclass[aps, pre, notitlepage, superscriptaddress, twocolumn,
               a4paper, floatfix]{revtex4-1}
\usepackage[utf8]{inputenc}
\usepackage{epsfig}
\usepackage[T1]{fontenc}
\usepackage{t1enc}
\usepackage{graphicx}
\usepackage{amssymb}
\usepackage{amsmath}
\usepackage{relsize}
\usepackage[caption=false]{subfig}
\usepackage{color}

\usepackage[normalem]{ulem}

\usepackage{mathtools}
\def\multiset#1#2{\ensuremath{\left(\kern-.3em\left(\genfrac{}{}{0pt}{}{#1}{#2}\right)\kern-.3em\right)}}

\begin{document}

\title{History effects in the sedimentation of light aerosols in turbulence: the case of marine snow}

\author{Ksenia Guseva}
\email{ksenia.guseva@uni-oldenburg.de}
\affiliation{Theoretical Physics/Complex Systems, ICBM, University of Oldenburg, 26129 Oldenburg, Germany}

\author{Anton Daitche}
\affiliation{Theoretical Physics/Complex Systems, ICBM, University of Oldenburg, 26129 Oldenburg, Germany}

\author{Ulrike Feudel}
\email{ulrike.feudel@uni-oldenburg.de}
\affiliation{Theoretical Physics/Complex Systems, ICBM, University of Oldenburg, 26129 Oldenburg, Germany}
\affiliation{MTA-ELTE Theoretical Physics Research Group, E\"otv\"os University,  P\'azm\'any P. s. 1/A,  H-1117, Budapest, Hungary}

\author{Tam\'as T\'el}
\email{tel@general.elte.hu}

\affiliation{MTA-ELTE Theoretical Physics Research Group, E\"otv\"os University,  P\'azm\'any P. s. 1/A,  H-1117, Budapest, Hungary}
\affiliation{Institute for Theoretical Physics E\"otv\"os University,  P\'azm\'any P. s. 1/A,  H-1117, Budapest, Hungary}


\begin{abstract}
  We analyze the effect of the Basset history force on the sedimentation of
  nearly neutrally buoyant particles, exemplified by marine snow, in a
  three-dimensional turbulent flow.  Particles are characterized by Stokes
  numbers much smaller than unity, and still water settling velocities, measured
  in units of the Kolmogorov velocity, of order one.  The presence of the
  history force in the Maxey-Riley equation leads to individual trajectories
  which differ strongly from the dynamics of both inertial particles without
  this force, and ideal settling tracers.  When considering, however, a large
  ensemble of particles, the statistical properties of all three dynamics become
  more similar. The main effect of the history force is a rather slow, power-law
  type convergence to an asymptotic settling velocity of the center of mass,
  which is found numerically to be the settling velocity in still fluid. The
  spatial extension of the ensemble grows diffusively after an initial ballistic
  growth lasting up to ca. one large eddy turnover time.  We demonstrate that
  the settling of the center of mass for such light aggregates is best
  approximated by the settling dynamics in still fluid found with the history
  force, on top of which fluctuations appear which follow very closely those of
  the turbulent velocity field.
\end{abstract}

\maketitle

\section{Introduction}

There is an increasing evidence, both theoretical and experimental, pointing out
the relevance of memory effects in the advection of inertial particles (see
e.g. \cite{tanga_dynamics_1994, belmonte_monotone_2001, daitche_memory_2011,
  guseva_influence_2013, daitche_memory_2014, farazmand_maxeyriley_2015,
  mordant_velocity_2000, candelier_effect_2004, vojir_effect_1994,
  langlois_asymptotic_2015}). Several further studies concerning these effects
in turbulence are reviewed in \cite{daitche_role_2015}. The equations of motion
for small spherical inertial particles were formulated by Maxey and
Riley~\cite{maxey_equation_1983} and Gatignol \cite{gatignol_faxen_1983} with
corrections by Auton et al.~\cite{auton_force_1988} and are of
integro-differential type in their full form. They contain an integral term
which accounts for the diffusion of vorticity around the particle throughout its
entire history.  This integral term is called the history (or Basset)
force~\cite{basset_treatise_1888}, and it has become clear by now that the often
used approximation in which this term is neglected is improper, and the full
Maxey-Riley equation should be considered ~\cite{tanga_dynamics_1994,
  belmonte_monotone_2001, daitche_memory_2011, guseva_influence_2013,
  daitche_memory_2014, farazmand_maxeyriley_2015, langlois_asymptotic_2015,
  mordant_velocity_2000, candelier_effect_2004, daitche_memory_2011,
  van_hinsberg_efficient_2011, vojir_effect_1994, daitche_advection_2013,
  daitche_role_2015}.

In this work, we analyze the effect of the history force on sedimenting
particles in turbulence in the presence of gravity. Previous efforts to
understand the importance of the history force in the presence of gravity in
smooth flows are due to Mordant and Pinton~\cite{mordant_velocity_2000} and to
Lohse and coworkers~\cite{toegel_viscosity_2006, garbin_history_2009} who also
carried out experiments. Their studies, however, concentrated on free
sedimentation, that is on the particle motion in a fluid at rest, and on bubble
dynamics in a standing wave, respectively. More recent papers investigate the
problem in a stationary~\cite{bergougnoux_motion_2014} and periodically changing
cellular flow~\cite{guseva_influence_2013}.  The sedimentation problem in
turbulent flows is considered up to now -- to our knowledge -- only in
stratified turbulence~\cite{aartrijk_vertical_2010} and for the plankton
problem~\cite{olivieri_analysis_2013}.

The motivation for our particular range of parameters comes from recent studies
of marine ecosystems which emphasize the importance of marine snow. Marine snow
plays a central role in the carbon cycle~\cite{mann_dynamics_2005,
  de_la_rocha_factors_2007,riley_relative_2012}, and its formation is mainly due
to physical aggregation, a consequence of particle-flow
interactions. Sedimentation of marine snow is considered to account for a large
fraction of carbon sequestration into the deep ocean~\cite{stone_invisible_2010,
  passow_biological_2012} this net ocean sequestration flux is estimated to
reach $\sim 10^{15}\; g\;Carbon/yr$~\cite{passow_biological_2012}. The physical
and biological properties of marine snow aggregates make it rather difficult to
estimate their sinking velocity. The techniques employed to evaluate settling
velocities vary across different measurements~\cite{mcdonnell_variability_2010,
  petrik_aggregates_2013}, and the results are difficult to compare due to the
variation in density and sizes of the used aggregates. An additional difficulty
for in situ experiments is the fact that turbulent kinetic energy varies with
depth~\cite{gargett_ocean_1989}. While some laboratory experiments with grid
generated turbulence \cite{murray_settling_1970,tooby_motion_1977} and in situ
measurements \cite{alldredge_situ_1988} find indications for a retarded settling
in situ compared to laboratory measurements in still water, other observations
in coastal areas \cite{shanks_abundance_2002} and in the laboratory using
Couette devices \cite{ruiz_turbulence_2004} report an enhancement of the sinking
speed in turbulence.

Marine snow particles contain organic and inorganic components as primary
particles which stick together in a fractal-like structure possessing a
relatively high porosity. This fact has been taken into account in concepts
working with an effective density~\footnote{The effective density is not easy to
  determine directly, but is often inferred from measured settling velocities
  based on the assumption of the validity of the Stokes law or of a modification
  thereof.}~\cite{kranenburg_fractal_1994, winterwerp_simple_1998}, an effective
diameter~\cite{logan_fractal_1990} or a modified Stokes
law~\cite{khelifa_models_2006} of the aggregates. Moreover, sinking marine
aggregates undergo changes in size and density due to aggregation and
fragmentation processes influencing the settling of them
~\cite{jackson_role_2005}.

\begin{figure}[h!]
\centering
\includegraphics[scale=.85]{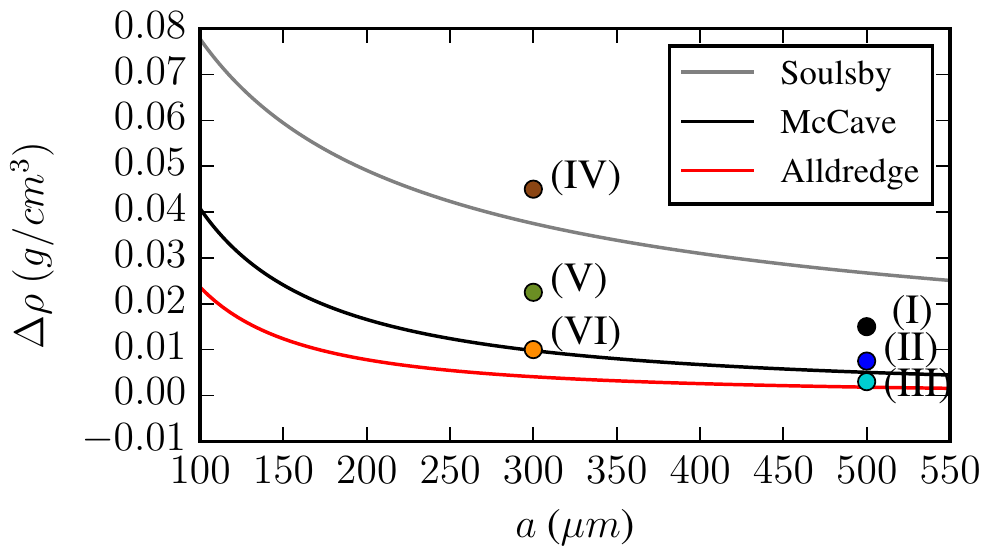}
\caption{Representation of the chosen parameters (see Table I) on the particle
  radius--excess density plane. The two lower curves represent the relationship
  between the effective excess density $\Delta \rho$ and the effective radius
  $a$ for aggregates with predominantly organic
  composition~\cite{mccave_size_1984, alldredge_situ_1988}, as expeceted for the
  open ocean, while the uppermost curve shows this relationship for aggregates
  from coastal areas and estuaries~\cite{soulsby_settling_2013}, containing a
  large fraction of inorganic components.}\label{fig:dens}
\end{figure}

Those biological properties are difficult to take into account when modeling the
sinking of marine aggregates as inertial particles using the Maxey-Riley
equation. The fractal shape can be taken into account by means of an effective
density~\cite{zahnow_particle-based_2011}, but this has been studied so far only
neglecting the history force.  Since the latter has not yet been formulated for
more complicated objects than spheres, the study presented here will work
exclusively with spherical particles the properties of which are based on the
effective aggregate diameters and effective densities given in the literature.
The effective densities of marine aggregates are usually very close to the water
density and a general property is that larger aggregates have smaller density
than small ones.  The relationship between the size (average effective radius
$a$) and effective excess density $\Delta \rho$ between particle and fluid
($\Delta \rho = \rho_p - \rho_f$) for aggregates rich in biological components
was proposed by {\it McCave et al}~\cite{mccave_size_1984} to be
$\Delta \rho \propto a^{-1.3}$ , and this relation was used to fit the
experimental results~\cite{tambo_physical_1979}, see middle curve in
Fig.~\ref{fig:dens}. This relation is close to the one obtained from in situ
measurements from Santa Barbara Channel by { \it Alldredge et
  al}~\cite{alldredge_situ_1988}, $\Delta \rho \propto a^{-1.6}$ for marine snow
characterized by $a > 250 \; \mu m$, also with predominately organic composition
(lowest curve Fig.~\ref{fig:dens}). On the other hand, in estuaries and coastal
regions aggregate composition includes more inorganic components
~\cite{simon_microbial_2002}, therefore, they are smaller and slightly denser
than the ones formed in the ocean. The corresponding effective size effective
density relationship was studied by {\it Soulsby et
  al}~\cite{soulsby_settling_2013}, and assumes $\Delta \rho \propto a^{-0.66}$
(see uppermost curve in Fig.~\ref{fig:dens}).  For a review
see~\cite{winterwerp_physical_2011}. Typical velocities in the ocean's upper
layer are strongly dependent on the wind and can reach up to $0.5$
m$/$s~\cite{oakey_dissipation_1982}. The turbulent kinetic energy $\epsilon$
typical for the open ocean is $\epsilon =10^{-6}$
m$^2/$s$^3$~\cite{mann_dynamics_2005, kirboe_planktivorous_1995}, which sets the
size of the smallest possible eddies, the Kolmogorov length $\eta$ to be
$\sim10^{-3}$ m.  The size of aggregates (macroaggregates) varies from $0.1$ to
less than $1$ mm~\cite{de_la_rocha_factors_2007, bartholoma_suspended_2009},
however the average aggregate size is always at most $\eta/2$ according to
~\cite{de_la_rocha_factors_2007, bartholoma_suspended_2009}, though the
relationship between the average aggregate size and the turbulent kinetic energy
in the ocean is not well established due to the difficulty of in situ
measurements.

Since we are interested in the effect of the history force we are confined to a
certain, yet realistic, set of parameters for size and density of our marine
snow particles which maximize the impact of the history force. On the one hand
we need Stokes numbers that are not too small for the history term to play an
important role. On the other hand, having a small Stokes number also decreases
the impact of preferential concentrations. To study the problem, we select six
density-size pairs typical of marine snow. The radii are 0.5 and 0.3 mm, since
the strongest impact of the history force is expected at the largest sizes,
largest possible Stokes numbers~\cite{daitche_role_2015}. To both of these sizes
we assign three different densities, see Fig~\ref{fig:dens}. We also display the
results of the relationship of the excess density $\Delta \rho$ and particle
diameter $a$ for open ocean~\cite{soulsby_settling_2013} and coastal
areas~\cite{mccave_size_1984}.  The sizes and the flow set the Stokes numbers
(see Eq.(\ref{St}) below) which take the values $St=0.083$ and $St=0.03$,
respectively. The parameters characterizing the six cases are summarized in
Table I.

\begin{table*}[t]
        \begin{center}
                \begin{tabular}{| p{1.5cm} ||  p{2.5cm} | p{2cm} | p{1.4cm} | p{1.cm} | p{1.cm} | p{2.cm} |p{3cm} | }
                  \hline
                  \centering case &\centering $\Delta \rho$ (g/cm$^3$) & $\beta$ &\centering  $a$ (m) &\centering  $St$ & $W$ &  $Re_z^{*}$ \\
                  \hline
                  \centering (I) & \centering 0.015 & 0.9900 &   &  &  8.21 & 4.1 \\
                  \cline{1-3} \cline{6-7}
                  \centering (II) & \centering 0.0075 & 0.9950 & \centering $5 \cdot 10^{-4}$ & \centering 0.083 & 4.12 & 2 \\
                  \cline{1-3} \cline{6-7}
                  \centering (III) &  \centering 0.003 & 0.9980&      &    & 1.65 & 0.8\\
                  \hline
                  \centering (IV) &  \centering 0.05 & 0.9677 &    &  &    9.67 & 2.9\\
                  \cline{1-3} \cline{6-7}
                  \centering (V) &  \centering 0.025 & 0.9836 & \centering $3 \cdot 10^{-4}$ &\centering 0.03 & 4.91 & 1.5 \\
                  \cline{1-3} \cline{6-7}
                  \centering (VI) &  \centering 0.01 & 0.9934 &    &  & 1.98 & 0.6\\
                  \hline
                \end{tabular}\caption{Parameters for six representative cases of marine aggregates in the ocean ((I), (II), (III)) and coastal areas ((IV), (V), (VI)).
                  Parameters $\beta$, $St$, $W$ and $Re^*_z$ are defined in equations
                  (\ref{beta}), (\ref{St}), (\ref{W}), and (\ref{Re*}), respectively,
                  and turbulence data are taken from Table II. }
        \end{center}
\end{table*}

As a preliminary qualitative analysis, let us concentrate here on the particle
Reynolds number $Re_p$ that should not exceed a limit. It should be below or of
the order of unity for the Stokesian drag to be valid, at least in a good
approximation. Since gravity breaks the isotropy of the advection problem by
preferring the vertical ($z$) direction, and particles are nearly neutrally
bouyant, it is worth defining a vertical and a horizontal particle Reynolds
number for spheres of radius $a$ and of typical slip velocity $\vec{v}-\vec{u}$
relative to the fluid
\begin{equation}
Re_z^*=\frac{a |v_z-u_z|}{\nu}, \qquad Re_h^*=\frac{a |\vec{v_h}-\vec{u_h}|}{\nu},
\label{Re*}
\end{equation}
where index $h$ refers to the horizontal component, and $\nu$ is the fluid's
kinematic viscosity.  The corresponding usual particle Reynolds number $Re_p$
follows from the identity
$Re_p^2=Re_z^2+Re_h^2$. 
The order of magnitude of the vertical slip velocity is the settling velocity in
still water which we write as $W_{\text{settling}}=W u_\eta$, where $W$ is the
dimensionless settling velocity taken in units of the Kolmogorov velocity
$u_\eta$ (all results for turbulent advection will be given in Kolmogorov
units). Measuring the particle radius in Kolmogorov length, $\eta$, we find
\begin{equation}
Re_z^*=\frac{a}{\eta} \frac{W u_\eta \eta}{\nu}= W \frac{a}{\eta},
\label{Re*z}
\end{equation}
since the fluid Reynolds number on the Kolmogorov scale $u_\eta \eta/\nu$ is by
definition unity.  The horizontal slip velocity is expected to vanish with the
Stokes number. Therefore, the horizontal slip velocity should be proportional to
$St u_\eta$. Taking the proportionality factor to be unity, we find for the
horizontal Reynolds number in an analogous manner the estimate
\begin{equation}
Re_h^*=St \frac{a}{\eta}.
\label{Re*h}
\end{equation}
Since $St \ll W$ (see Table I.), we find that $Re_p \approx Re_z$. We think that
this is a central property of marine snow sedimentation, which expresses that
these particles behave horizontally as nearly neutrally bouyant, but they
sediment with a speed comparable to that of the small scale fluctuations of the
fluid ($u_{\eta}$), i.e. they are not neutrally bouyant from the point of view
of the vertical dynamics. We shall in fact see that the instantaneous particle
Reynolds numbers converge in time towards $Re_z^*$. The characteristic numbers
$Re^*_z$ are also indicated in Table I.

The paper is organized as follows: In Sec.~\ref{sec:overview} we present an
overview of the equation of motion with the history force. Next, we recall an
infinite series solution of it in still fluid and find a simple analytic
approximation to be valid after relatively short times, presented in
Sec.~\ref{sec:still}. Then we summarize the approach used to compute the history
force, and to generate the turbulent velocity field in Sections~\ref{sec:flow}.
In Sec.~\ref{sec:positions} our numerical results concerning the sedimentation
dynamics in space are summarized. In section~\ref{sec:velocity} we turn to
results on velocities and accelerations.  Sec.~\ref{sec:Faxen} is devoted to
estimating the relevance of the Fax\'en corrections. Our final conclusions are
given in Sec.~\ref{sec:conclusion}.

\section{Equation of motion and notations}\label{sec:overview}

We analyze the advection of spherical, rigid particles with a small particle
Reynolds number in an incompressible and viscous fluid. The Lagrangian
trajectories of such particles are evaluated according to the Maxey-Riley
equation~\cite{maxey_equation_1983,gatignol_faxen_1983}, including the
corrections by Auton and coworkers~\cite{auton_force_1988}. In the full
Maxey-Riley picture one describes the dimensionless evolution of the particle
position $\vec{x}(t)$ and velocity $\vec{v}(t) = d\vec{x}/dt$ in a flow field
$\vec{u}(\vec{x}, t)$. Without Fax\'en corrections the equation of motion reads
as

\begin{equation}
\frac{d\vec{v}}{dt} = \frac{1}{St}(\vec{u} -\vec{v})  + \frac{W}{St}\vec{n} + \beta\frac{D\vec{u}}{Dt} - \sqrt{\frac{3 \beta}{\pi St}}\int_0^t \frac{\frac{d(\vec{v} - \vec{u})}{d\tau}}{\sqrt{t-\tau}}d\tau,
\label{eq:MR}
\end{equation}
where $\vec{n}$ is the vertical unit vector pointing downwards. This form of the
equation holds when the particle is initialized at time zero with a velocity
coinciding with that of the fluid, zero initial slip velocity.  We have to
distinguish the full derivative along a fluid element and a particle trajectory,
given by
\begin{equation*}
\frac{D}{Dt} = \frac{\partial }{\partial t} + \vec{u} \cdot \nabla \qquad \text{and} \qquad \frac{d }{dt} = \frac{\partial}{\partial t} + \vec{v} \cdot \nabla,
\end{equation*}
respectively. The velocity of the particle changes due to the action of
different forces. The forces in (\ref{eq:MR}) represent from left to right: the
Stokes drag, the gravity, the pressure force (which accounts for the force felt
by a fluid element together with the added mass force), and lastly the Basset
history force. The equation is written in dimensionless form, rescaled by the
Kolmogorov time $\tau$ and the Kolmogorov length scale $\eta$ of the flow
($u_\eta=\eta/\tau_{\eta})$. The ratio
\begin{equation}
\beta =\frac{3\rho_f}{ \rho_f + 2\rho_p} =\frac{3\rho_f}{3 \rho_f + 2\Delta \rho}
\label{beta}
\end{equation}
characterizes the excess density of the particle $\Delta \rho$ and the density
of the fluid $\rho_f$. For aerosols $\beta<1$~\footnote{For small excess densities $\Delta \rho$, characteristic to our cases, $\beta=1-\frac{2}{3}\frac{\Delta \rho}{\rho_f}$, as follows from Eq.(~\ref{beta}) for small $\Delta \rho/\rho_f$.}.

Another dimensionless parameter in Eq. (\ref{eq:MR}) is the Stokes number
\begin{equation}
St =  \frac{a^2}{3 \nu \beta \tau_\eta}=\frac{\tau_p}{\tau_\eta},
\label{St}
\end{equation}
which is the ratio of the particles' relaxation time $\tau_p$ due to kinematic
viscosity $\nu$ of the fluid to the Kolmogorov time.

Additionally, parameter $W$ governs the dimensionless settling velocity in still
fluid.
It can be written as
\begin{equation}
  W = St (\beta-1) \frac{g \eta}{u^2_\eta},
\label{W}
\end{equation}
where the last factor corresponds to the reciprocal of a turbulent Froude
number.  It is to be emphasized that $W$ cannot be varied freely: an ad-hoc
choice of $W$ to a given $St$ could imply that, for a fixed density, the flow
and/or the gravity $g$ are changed.  The $W$ values given in Table I., used by
us, are the ones which follow from the particle properties and the
characteristics of our turbulent flow.  Anyhow, in sedimentation the role of the
dimensionless settling velocity might be more relevant than that of the Stokes
number.

We shall compare the Maxey-Riley equation (Eq.(\ref{eq:MR})) to the
approximation which does not take into account the history force,
\begin{equation}
  \frac{d\vec{v}}{dt} = \frac{1}{St}(\vec{u} -\vec{v} + W\vec{n}) + \beta\frac{D\vec{u}}{Dt},
\label{eq:A1}
\end{equation}
often called the advective equation of inertial particles. We emphasize that
Eq.(\ref{eq:A1}) does not follow from any approximation of the Maxey-Riley
equation for our sets of particle parameters, its use is motivated by mere
numerical convenience.

We also carry out simulations with the equation
\begin{equation}
 \vec{v} = \vec{u} +W\vec{n},
\label{eq:id}
\end{equation}
valid for ideal non-inertial particles. Note that this case arises when
$St \rightarrow 0$, and is the limit of both equations (\ref{eq:MR}) and
(\ref{eq:A1}), with different convergence properties, of course.

\section{Settling in still fluid}\label{sec:still}

The exact solution for the settling in a still fluid $(\vec{u} = 0)$ was worked
out by Belmonte and coworkers~\cite{belmonte_monotone_2001}. In this case a
natural velocity unit is the settling velocity $W_{\text{settling}}$, and time
can be measured in units of the particle relaxation time $\tau_p$.  In these
units, the dimensionless vertical velocity $v'_z(t')$ in dimensionless time $t'$
can be expressed in terms of complementary error functions $erfc$. With zero
initial velocity it reads in our notation as

\begin{multline}
v'_z(t') =  1  + \frac{\sqrt 3 \beta}{ \alpha_1 -\alpha_2}\bigg[\frac{e^{\alpha_1 t'} \text{Erfc}\left( \sqrt{\alpha_1 t'}\right)}{\sqrt{\alpha_1}} 
\\ - \frac{e^{\alpha_2 t'} \text{Erfc}\left( \sqrt{\alpha_2 t'}\right)}{\sqrt{\alpha_2}} \bigg]
\label{eq:free}
\end{multline}
where $\alpha_1$, $\alpha_2$ are the roots of the quadratic equation
$\alpha^2 + (2-3 \beta) \alpha + 1=0$
depending only on the density via parameter $\beta$.

By keeping only the leading terms of the power law expansion of the function
$e^{u}$Erfc$(\sqrt{u})$ for large $u$ (long times $t'$), we find
\begin{equation}
v'_z(t') = 1 - \sqrt{\frac{3\beta}{\pi t'}} \left(1 - \frac{(3\beta - 2)}{2 t'}\right).
\label{eq:solution_app}
\end{equation}
This form turns out to provide a rather accurate approximation for $t' > 2$, and
even by neglecting the second term in the parenthesis it is very close to the
exact solution for $t'> 22$.  Note that these forms do not depend on the particle
size since Stokes numbers can only be defined in a moving fluid.  Whether the
particle Reynolds number $Re_p$ remains small, i.e. whether equation
(\ref{eq:MR}) remains valid during the entire free fall, should be checked a
posteriori in the knowledge of the dimensional settling velocity, the particle
size and the fluid's kinematic viscosity.

For comparison, we mention that the solution of the widely used inertial
dynamics equation (Eq.(\ref{eq:A1})) provides for the same problem a linear
differential equation whose solution is with the same zero initial condition,
and in the same units:
\begin{equation}
v'_z(t') = 1 - e^{-t'}.
\label{eq:solution_appi}
\end{equation}
This solution is of completely different character.

\begin{figure}[h!]
\centering
\includegraphics[scale=.85]{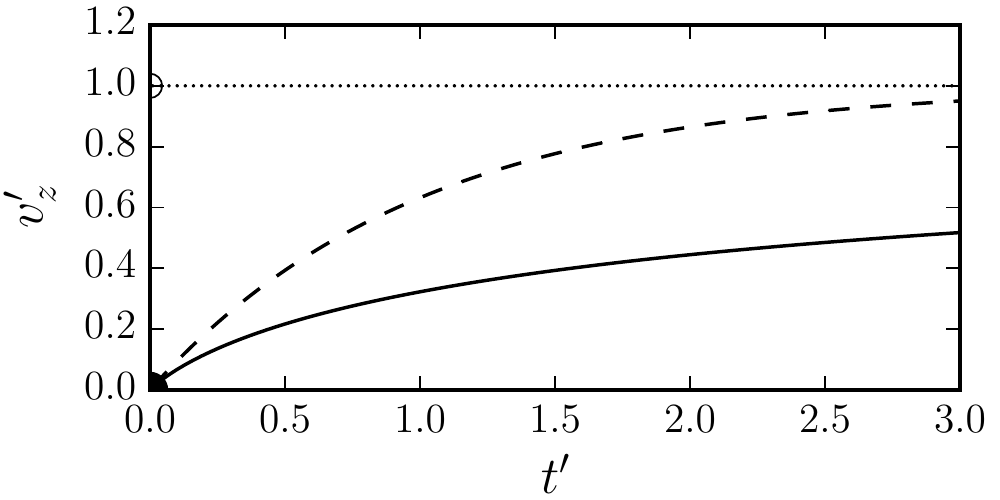}
\caption{Short-term behavior of the settling in still fluid ($\vec{u}=0$) in the
  different dynamics investigated. With memory (\ref{eq:free}): continuous line;
  without memory (\ref{eq:solution_appi}): dashed line, and non-inertial
  particles ($v'_z=1$ for $t'>0$, as follows from (\ref{eq:id})): dotted line.
  Note the rather different velocities predicted for any time instant. Time and
  velocity are measured in units of $\tau_p$ and $W_{\text{settling}}$,
  respectively.}\label{fig:stillfluid}
\end{figure}

The solution of the ideal tracer problem (Eq.(\ref{eq:id})) is that the particle
velocity jumps immediately from $0$ to unity and remains there forever. Note
that this behavior follows from both formulas (\ref{eq:solution_app}) and
(\ref{eq:solution_appi}) in the limit of $\tau_p \rightarrow 0$, which is
equivalent to taking $t' \rightarrow \infty$ in these
expressions. Eq. (\ref{eq:solution_appi}), however, does not follow as any limit of (\ref{eq:free}). Fig.~\ref{fig:stillfluid} provides a comparison of these
different dynamics.

\begin{table*}[t]
\begin{center}
\begin{tabular}{| p{1.5cm}|p{1.5cm}|p{1.5cm}|p{1.5cm}|p{1.5cm}|p{1.5cm}|p{1.5cm}|p{1.5cm}|p{1.5cm}|p{1.5cm}|}
\hline
\centering $Re_{\lambda}$ & \centering $L_{{\rm box}}/\eta$ & \centering$L/\eta$ & \centering $\lambda/\eta$ & \centering$\Delta
x/\eta$ & \centering $T_{\mathrm{sim}}/\tau_{\eta}$ & \centering$T/\tau_{\eta}$ & \centering $\Delta t/\tau_
{\eta}$ & \centering $u_{\mathrm{rms}}/u_{\eta}$ & \centering $N^{3}$\tabularnewline

\hline
\centering $112$ & \centering $633$ & \centering $156$ & \centering $20.9$ & \centering $1.24$ & \centering $1020$ & \centering $29.0$ & \centering $0.015$ & \centering $5.39$ & \centering$512^{3}$\tabularnewline
\hline
\end{tabular}
\end{center}
\caption{\label{tab:sim-para}Parameters of the simulated turbulent flow: Taylor
Reynolds number $Re_{\lambda}=\lambda u_{\mathrm{rms}}/\nu$, size of
the periodic box $L_{{\rm box}}$, integral scale $L=u_{\mathrm{rms}}^{3}/\epsilon$,
Taylor microscale $\lambda=u_{\mathrm{rms}}\sqrt{15\nu/\epsilon}$,
size of a grid cell $\Delta x$, length of the whole simulation $T_{\mathrm{sim}}$,
large-eddy turnover time $T=L/u_{\mathrm{rms}}$, time step $\Delta t$,
root-mean-square of the velocity $u_{\mathrm{rms}}=\sqrt{\left\langle \vec{u}^
{2}\right\rangle /3}$,
number of grid points $N^{3}$. All dimensional quantities are given
in multiples of the corresponding Kolmogorov units.}
\end{table*}

\section{Turbulent flow and numerical Simulation}\label{sec:flow}

We consider here the case of particles moving in statistically homogeneous,
isotropic and stationary turbulence \cite{pope_turbulent_2000}. To this end we
solve the vorticity equation, which is equivalent to the incompressible
Naiver-Stokes equation, on a grid in a triply-periodic box of size
$L_{box}$. The energy is injected by a large scale forcing, see
\cite{daitche_role_2015}. For the integration of the flow we use a standard
dealiased Fourier-pseudo-spectral method \cite{canuto_spectral_1987,
  hou_computing_2007} with a third-order Runge-Kutta time-stepping scheme
\cite{shu_efficient_1988}. The values of Eulerian quantities, which are
available on a grid, are obtained at the particle positions through tricubic
interpolation. The characteristics of the turbulent flow and the simulation
parameters are depicted in Table \ref{tab:sim-para}.  Since the Kolmogorov scale
is $\eta=(\nu^3/\epsilon)^{1/4}$ \cite{pope_turbulent_2000}, a fixed value of it
can belong to any kinematic viscosity $\nu$ and mean energy dissipation
$\epsilon$, as long as the ratio $\nu^3/\epsilon$ is fixed.  For the particular
choice of $\eta=1$ mm, which we shall take as a typical value in our
estimations, one finds with the viscosity of water $\epsilon \sim 10^{-6}$
m$^2/$s$^3$.

The presence of the history integral in (\ref{eq:MR}) leads to two problems from
the numerical point of view. First, the singularity of the history kernel
impedes an accurate numerical solution. This problem can be solved by the use of
a specialized integration scheme \cite{daitche_advection_2013} which treats the
history force appropriately. This third order scheme has been adjusted for our
purposes, see \cite{daitche_role_2015} for details. Second, it is necessary to
recompute the history integral for every new time step. This leads to high
computational costs and a high demand for memory (to store the history of each
particle). This second problem is inherent to the dynamics with memory and, as a
consequence, limits us to a moderate number of particles. For each case of
particle parameters we simulated $N_p = 1.5\cdot10^{5}$ particles. The initial
particle positions have been chosen randomly and homogeneously distributed in
the triple-periodic box of size $L_{box}$ of the simulation; the initial
particle velocity is that of the fluid at the particle's position.

\section{Turbulence: Results on the position of particles}\label{sec:positions}

We start by comparing individual trajectories in the Maxey-Riley equation
(\ref{eq:MR}), in the inertial equation (\ref{eq:A1}) in which memory is
neglected and in the non-inertial dynamics (\ref{eq:id}).  Throughout the paper
we will use the following notation for particles following the different
dynamical equations (\ref{eq:MR}), (\ref{eq:A1}), and (\ref{eq:id}), and show
their corresponding curves with particular line types:
\begin{itemize}
\item particles {\em with memory}, continuous line, computed by (\ref{eq:MR}),
\item particles {\em without memory}, dashed line  computed by (\ref{eq:A1}), and
\item {\em non-inertial} particle, dotted line,  computed by (\ref{eq:id}),
\end{itemize}
respectively.

\begin{figure*}[t]
\centering
\includegraphics[scale=.75]{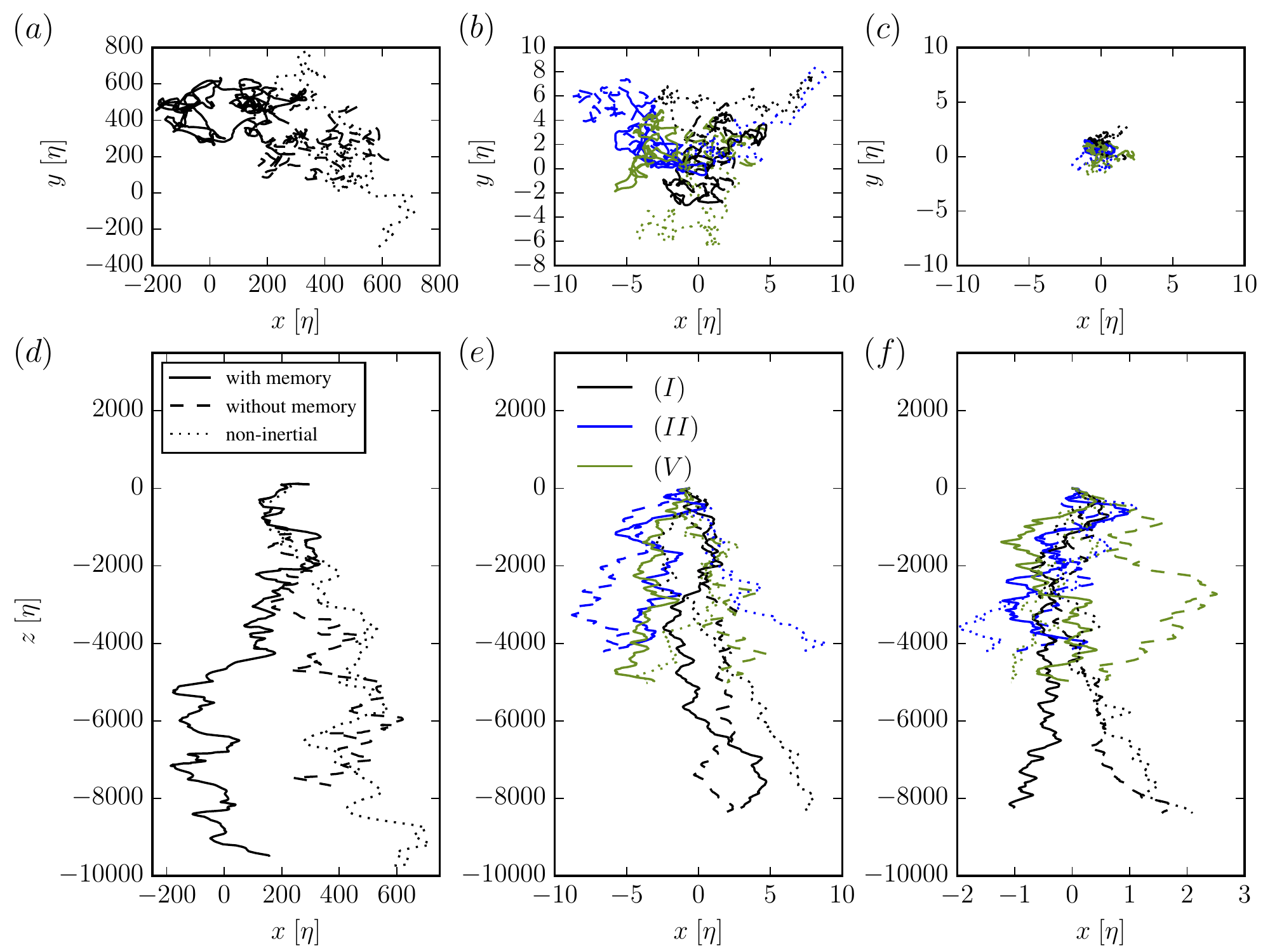}
\caption{(left) Individual trajectories of the same particle started with the
  same initial conditions (with zero slip velocity) with the parameter set of
  case (I), however, following three distinct equations of motion 
  ($x_0 = 293.16 \; \eta$, $y_0= 214.18\; \eta$, $z_0 = 105.71\; \eta$).
  (middle and right) Position of the center of mass of the ensemble containing
  $10 \%$ of our standard particle number ($N_p = 1.5 \cdot 10^4$), and the
  standard number $N_p = 1.5 \cdot 10^5$ of particles, respectively, for the
  case (I), (II), and (V) evolved with the different equations of motion up to
  $1020 \;\tau_{\eta}$. Here and in the following figures we use the convention
  that x and t denote the dimensional space and time, respectively, and the
  dimensions are given in parentheses.
  }\label{fig:traj_individual}
\end{figure*}

Trajectories with the same initial condition, but following these three distinct
dynamics, deviate from each other already after a short period of time. The
distance among these trajectories increases significantly with time in both the
horizontal and the vertical directions. This results in strong differences in
the predictions for the position of a particle, since the trajectories of
different dynamics can be thousand Kolmogorov lengths away from each other after
$500 \tau_\eta$ as Fig.~\ref{fig:traj_individual}a,d illustrates~\footnote{Given
  the same initial condition on the same computer (i.e. floating-point
  architecture) with a fixed time step, would not lead to a separation of two
  runs even in chaos with any chosen equation of motion. Note that the
  deviations generated by our three different dynamics can be additionally
  amplified by the chaotic nature of the particle dynamics. Here, were are
  unable to separate out this amplification.}.

Although there are strong differences for the predictions of the position of an
individual particle for these three dynamics Fig.~\ref{fig:traj_individual}d,
these differences are smaller when an ensemble is considered
Fig.~\ref{fig:traj_individual}e,f. For this analysis we initialize clouds of
particles, one with smaller and other with larger number of particles, with the
initial condition mentioned above, and evolve them according to our three
possible dynamics. The center of mass of each cloud also follows a distinct
trajectory, however the distances among the centers of mass do not grow as fast
as that of the trajectories of individual particles, see
Fig.~\ref{fig:traj_individual}b,c where the final horizontal difference is of a
few Kolmogorov lengths only. Moreover, in the $x, y$ planes (upper panels a--c),
it becomes clear that the total displacements are in rather different directions
with the different dynamics. The horizontal distances between the centers of
mass decrease with the number of particles, which can be considered as a
consequence of the law of large numbers. The total horizontal displacement in
all three dynamics is therefore expected to be zero in the large particle number
limit~\footnote{The limit of $N_p \rightarrow \infty$ can only be considered for
  mathematical convenience. In the case of finite size particles it is important
  to remain in the dilute limit, in order to avoid hydrodynamic interactions.}.

\begin{figure*}[t]
\centering
\includegraphics[scale=.5]{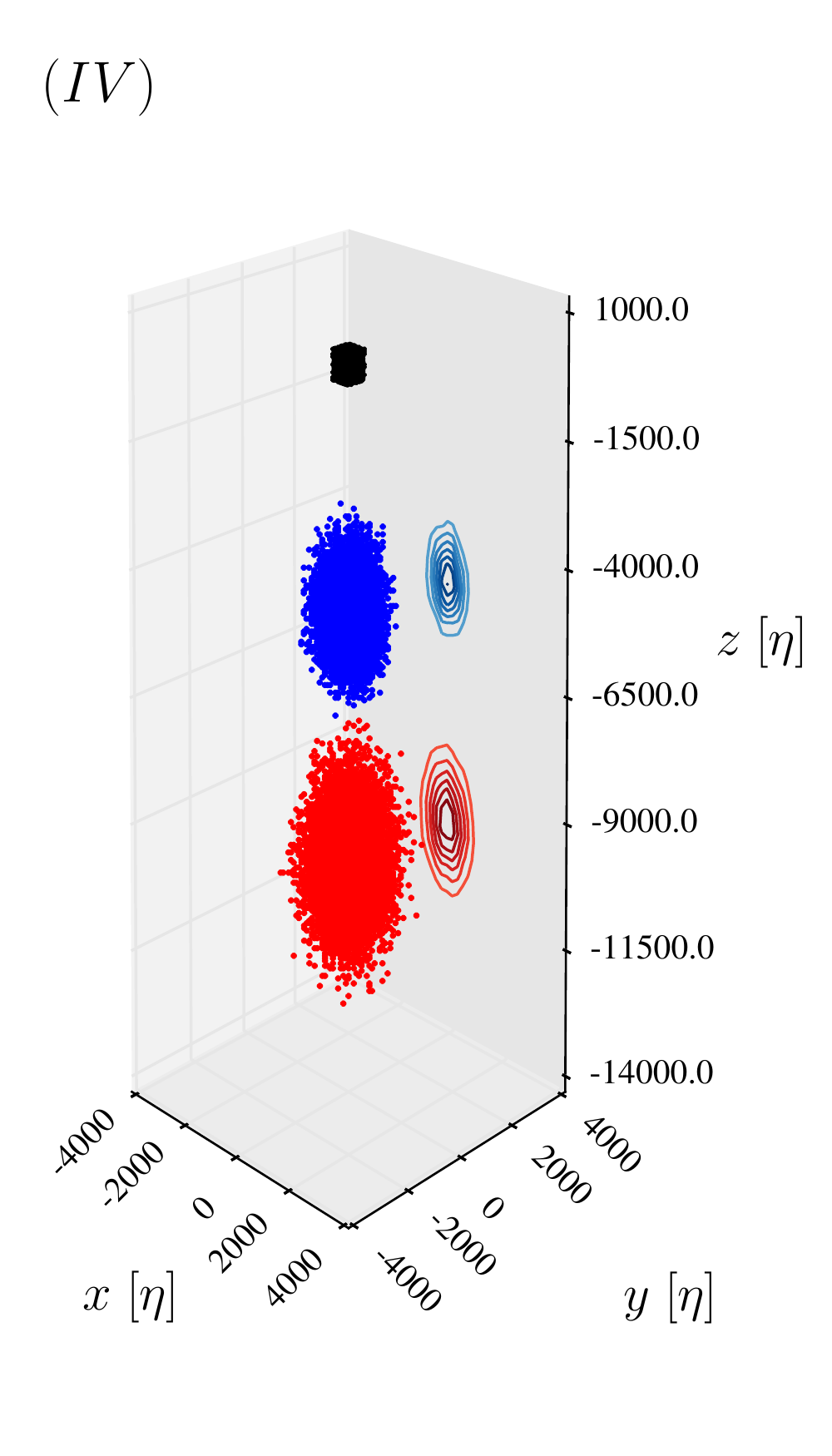}
\includegraphics[scale=.5]{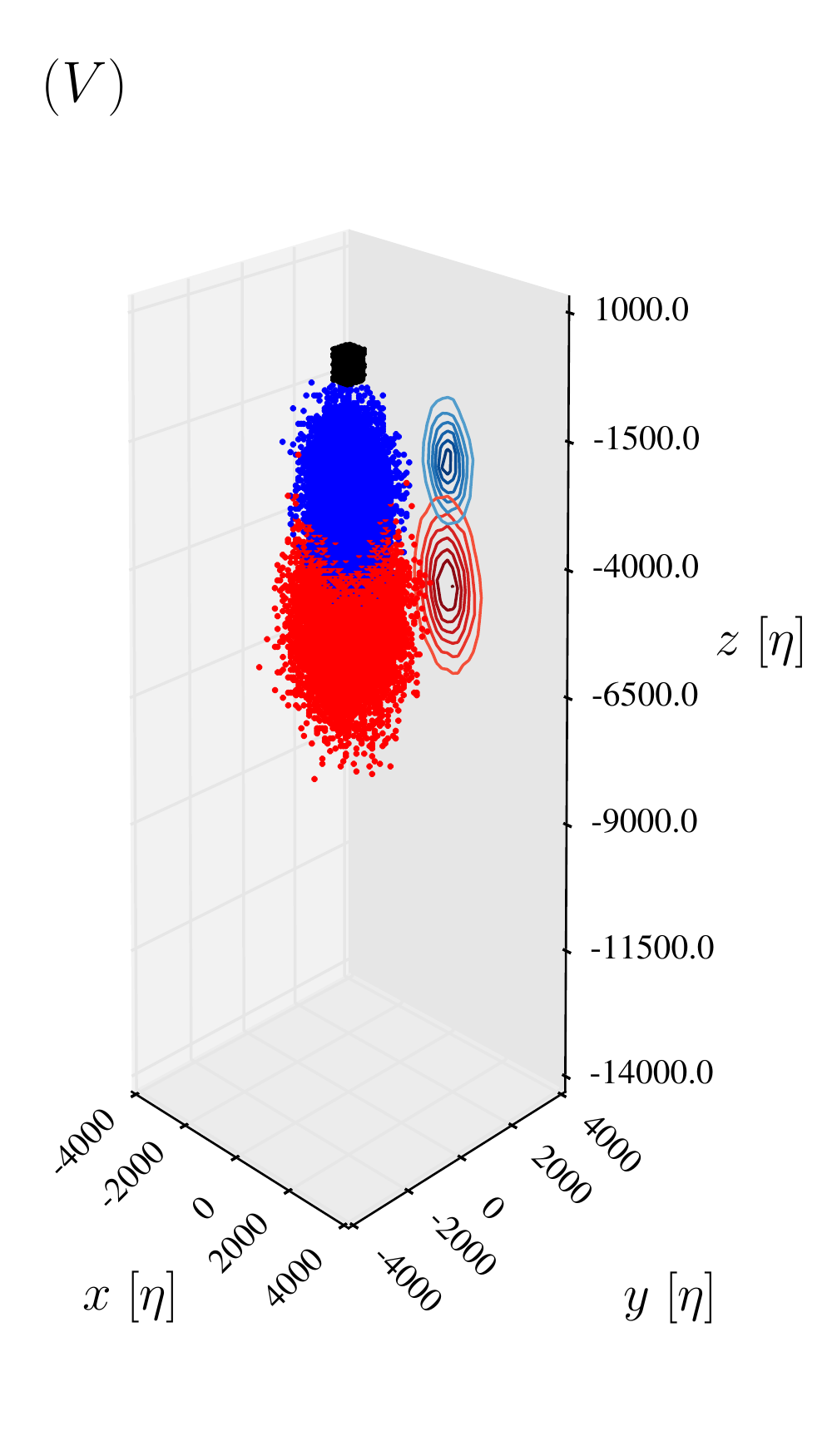}
\includegraphics[scale=.5]{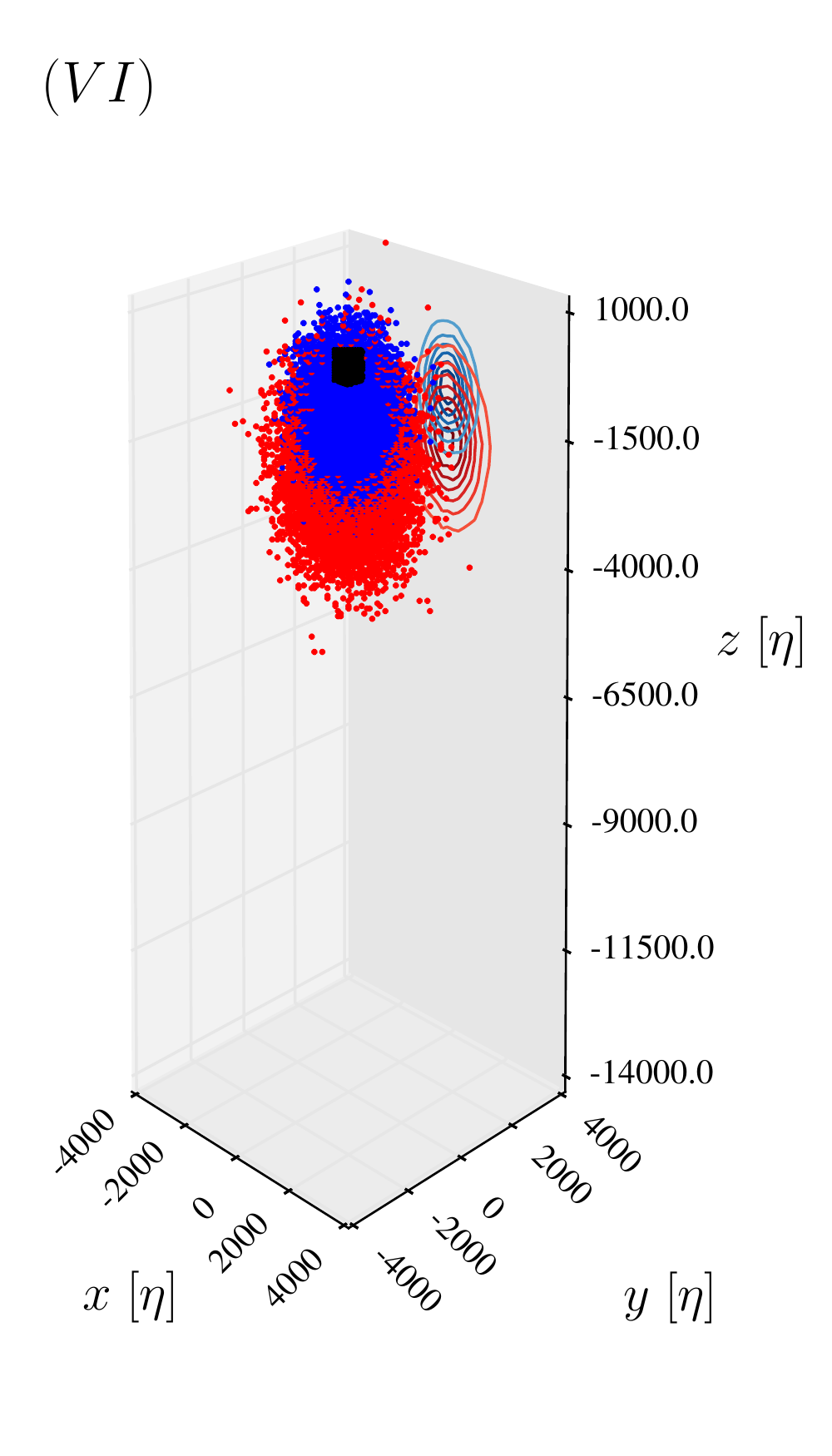}
\caption{Spatial distribution of the sedimenting particle ensembles
  for cases (IV), (V) and (VI) simulated with the Maxey-Riley equation
  (\ref{eq:MR}). Black, blue, and red dots represent the location of the
  particles at times $t=0$, $510 \; \tau_{\eta}$, and $1020 \; \tau_{\eta}$,
  respectively. A 2D histogram of the particle density is projected onto the
  $(y, z)$ plane, curves represent isolines of densities.}
\label{fig:ensemble}
\end{figure*}

After having seen the results for the center of mass of the particle ensembles,
we show in Fig.~\ref{fig:ensemble} their distribution in space at three
different time instants.  It is clear that with large settling velocities the
ensemble blobs are well separated after 500 time units, this separation
decreases, however, with $W$, and with the smallest settling velocity there is
hardly any separation, the blobs strongly overlap, and some points are even
above the cube of initial conditions after 1020 time units.  These results are
obtained in the presence of memory effects but we generated the corresponding
figures without memory and with non-inertial particles (governed by
Eqs.(\ref{eq:A1}) and (\ref{eq:id}), respectively), too. No difference can be
recognized by naked eyes. This is the first hint to the fact that inspite of the
difference in the individual and in the center of mass trajectories following
from the different dynamics, the statistical properties are quite similar. To
see the differences, more quantitative methods should be taken.

\begin{figure}[h]
\centering
\includegraphics[scale=0.7]{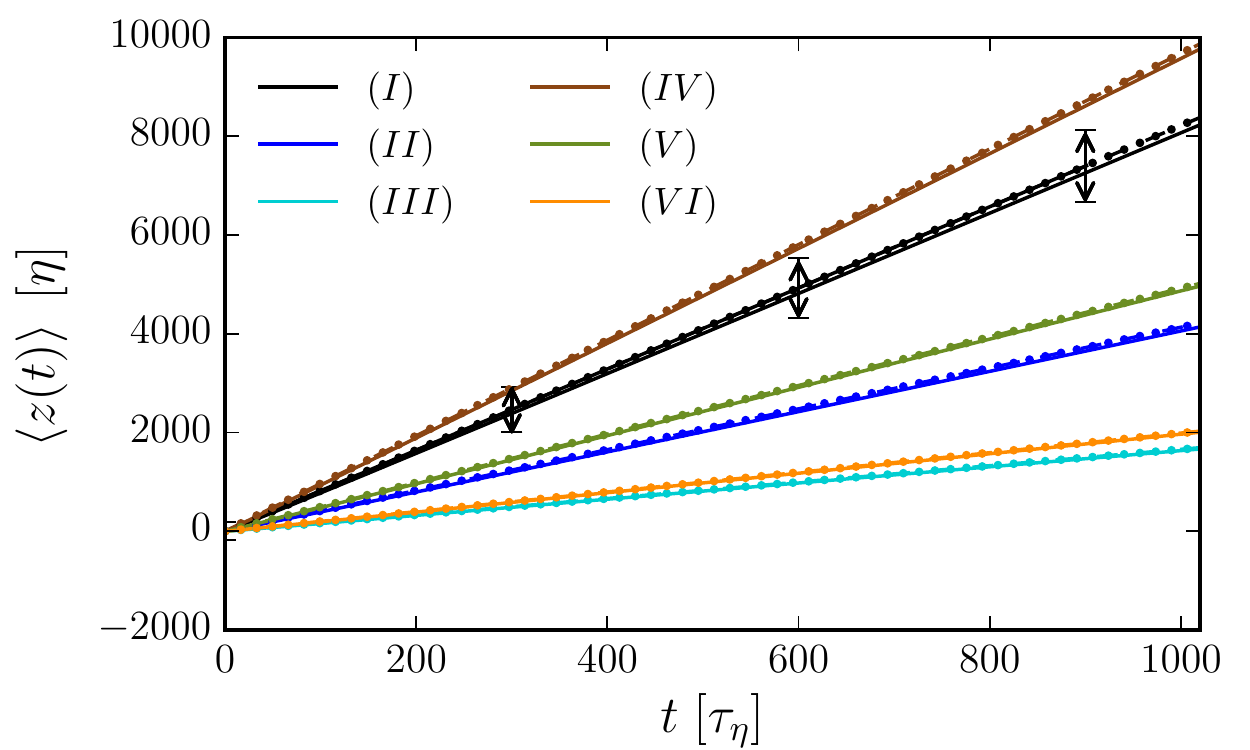}
\caption{Time-dependence of the $z$-coordinate of the center of mass in the
  different dynamics, distinguished by different line types, and for all six cases
  distinguished by different colors, dashed
  and dotted lines are hard to distinguished in this representation.
  The two-sided arrows indicate the typical spatial extension in $z$ dimensions
  of the ensemble for case (I) at the given instances.}\label{fig:z}
\end{figure}

In Figure~\ref{fig:z} we present a time-continuous plot of the $z$-coordinate of
the center of mass for the different cases with the three different dynamics.
Here differences become visible, and are on the order of a few hundred
Kolmogorov lengths at the end of the simulation. The difference is, however,
never larger than a few percent of the instantaneous value of
$\left<z\right>$. The long-term behavior is a roughly linear increase in all
cases, indicating a rather uniform settling.  Note that the graphs for the
largest excess density (case (I) and (IV)) are close to each other in spite of
the different Stokes numbers. The cases with intermediate and small excess
densities behave also similarly.

\begin{figure}[h]
\centering
\includegraphics[scale=.9]{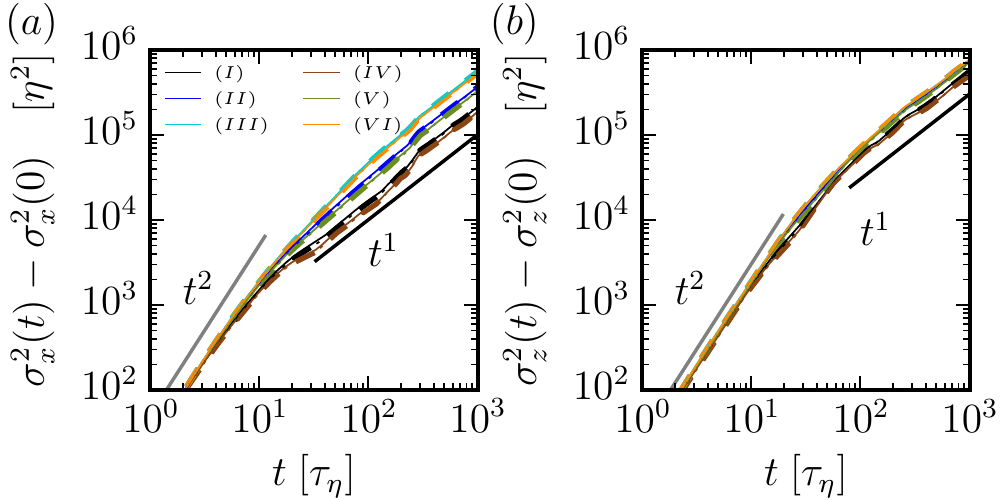}
\caption{Time-dependence of the variances in the horizontal ($x$, left panel)
  and vertical ($z$ right panel) directions in the different cases (coloring and
  line types as in as in the previous figure) on log-log scales. For clarity,
  the initial variance $\sigma_x(0)$ ($= 633/\sqrt{12}=183$ in dimensionless
  units) is subtracted. The graphs of different dynamics are overlaid and dashed
  and dotted lines are hard to distinguished in this representation.  The thin
  continuous black lines have slopes $2$ and $1$, respectively, to guide the
  eye. }\label{fig:sigma}
\end{figure}

As seen from Fig.~\ref{fig:ensemble}, the blob sizes are also important. To
monitor their time evolution, we determined the standard deviation $\sigma$
about the center of mass in the horizontal $x$ direction, and in the vertical
direction, as Fig.~\ref{fig:sigma} shows. The horizontal and the vertical
behavior are different, reflecting again that gravity prefers a certain
direction. In fact, the vertical extension of all the blobs is larger than the
horizontal one at any instant. The data are plotted on a log-log scale to
enlighten the appearance of power law behavior.  The two black straight lines
represent ballistic ($\sigma^2 \sim t^2$) and diffusive ($\sigma^2 \sim t$)
spreading. A crossover to the diffusive behavior can be observed at
$t \sim 20-30$ time units.  It is natural to understand that when the blobs are
large, the ensembles become subjected to a diffusive spreading by wandering
in-between the largest scale vortices.

In order to explain the ballistic behavior, we recall the theory of
Batchelor~\cite{batchelor_application_1950} for the separation of pairs of ideal
tracers in three-dimensional homogeneous isotropic turbulence.  This theory
claims that the mean square separation should grow as $t^2$ for times shorter
than a characteristic time $t_0$. For times larger than $t_0$ the famous
Richardson scaling~\cite{pope_turbulent_2000} should hold characterized by a
scaling proportional to $t^3$. This regimes extends, however, only up to the
time when the effect of the largest coherent structures becomes dominant,
i.e. up to the eddy turnover time $T$.  The characteristic time $t_0$ depends on
the initial spatial separation $\vec{r}_0$ between the two particles.  In
dimensional units $t_0=(\mid \vec{r}_0 \mid^2/\epsilon)^{1/3}$. Hence, for an
ensemble of particles with different initial distances no unique $t_0$ can be
found, so that only a typical $t_0$ can be estimated.

Although the original theory applies to ideal, i.e. non-settling tracers, it is
worth estimating $t_0$. For our initial ensemble a natural choice is the
variance of their positions in the initial cube of size $L_{box}$, what is
$\sqrt{3} L_{box}/\sqrt{12}=\sqrt{3} \sigma_x(0) \eta$, where $\sigma_x(0)$
denotes the dimensionless variance in $x$ direction, used in the plots of
Fig.~\ref{fig:sigma}. To estimate the dimensionless $t_0/\tau_\eta$ we replace
$\mid \vec{r}_0 \mid^2$ by $3 \sigma^2_x(0) \eta^2$ to find
$$
\frac{t_0}{\tau_\eta}=\left(\frac{3 \sigma^2_x(0) \eta^2}{\epsilon \tau^3_\eta} \right)^{1/3}=
(3 \sigma^2_x(0))^{1/3}=(3 \times 183^2)^{1/3}=46,
$$
where we used that $\tau_\eta=(\nu/\epsilon)^{1/2}$ and
$\eta=(\nu^3/\epsilon)^{1/4}$~\cite{pope_turbulent_2000}.  This value turns out
to be larger than the dimensionless turnover time, $T/\tau_\eta$ which is about
30 (see Table~\ref{tab:sim-para}). Thus, there is no possibility for seeing the
Richardson scaling due to the broad initial distribution of the particles.  The
anisotropy of the problem is reflected in the fact that the crossover to the
diffusive behavior occurs somewhat later in the vertical than in the
horizontal. We have verified that the evolution of an initially strongly
localized ensemble leads to Richardson behaviour (see Appendix A).

It is worth mentioning a related problem. Single particle dispersion
was investigated in stratified turbulence by van Aartrijk and Clercx
in the presence of the history force~\cite{aartrijk_vertical_2010},
but with larger typical excess densities and Stokes numbers compared
to ours.  They also found a crossover between ballistic and diffusive
spread, i.e. the history force did not change the exponents.

\section{Turbulence: Results on velocities and accelerations}\label{sec:velocity}

\begin{figure}[h]
\centering
\includegraphics[scale=0.55]{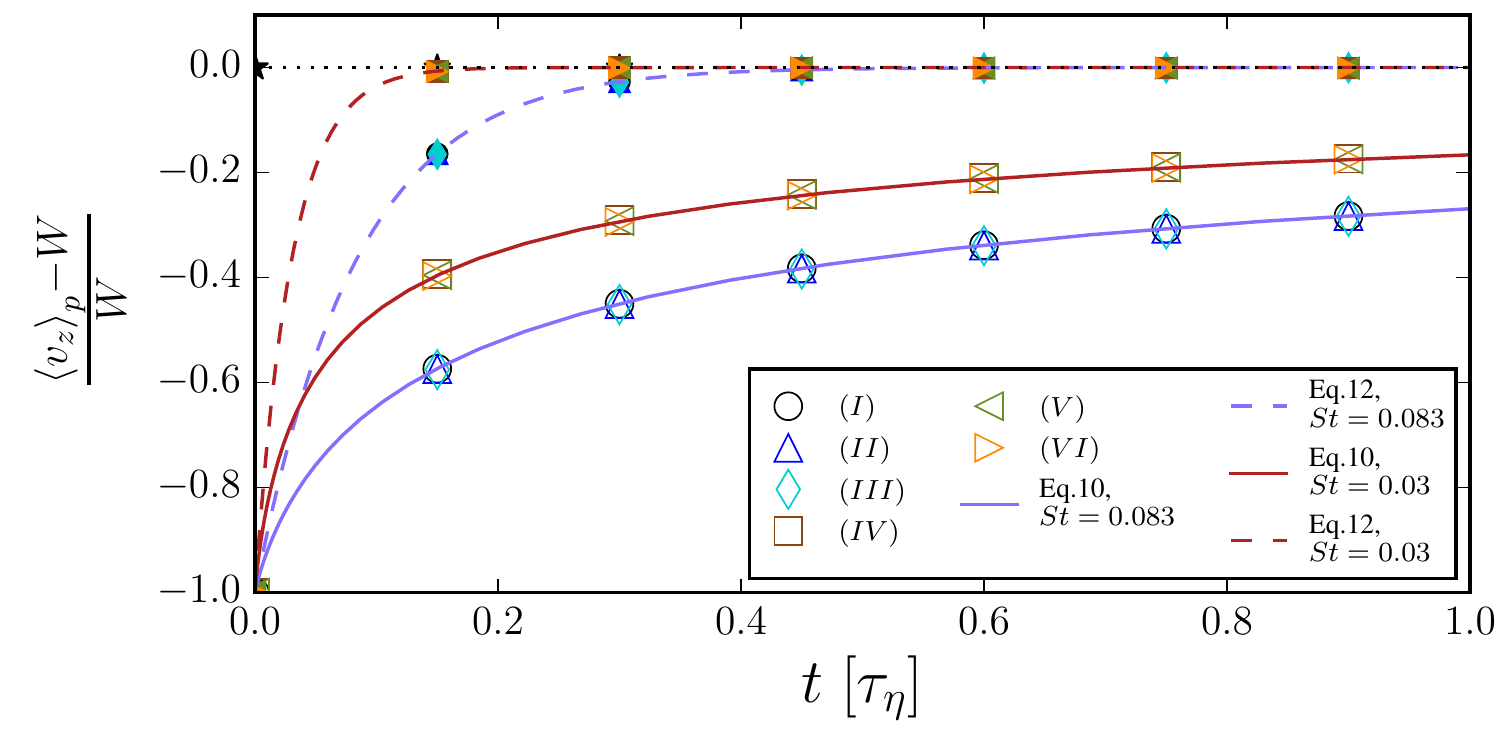}
\caption{Short-term behavior of the center of mass velocities expressed as
  $(\left<v_z\right>-W)/W$: empty symbols: with memory (Eq.~\ref{eq:MR}), and
  full symbols: without memory (Eq.~\ref{eq:A1}). Black stars indicate the
  results for non-inertial particles but only up to $t=0.3 \; \tau_{\eta}$ in
  order to avoid heavy overlap. Continuous (dashed) curve represents the still
  fluid result with memory (without memory) as expressed by (\ref{eq:free})
  ((\ref{eq:solution_appi})). }\label{fig:vel_short}
\end{figure}

Inertial effects are more pronounced in the velocity data characterizing the
ensemble.  An investigation of the short-term behavior, up to a single time unit
(one Kolmogorov time), indicates clearly that particles for which the history
force is neglected approach typically much faster the asymptotic settling
velocity than those for which the history force is taken into account. This can
very well be seen in Fig.~\ref{fig:vel_short} which exhibits the $z$-component
of the particle velocities averaged over the ensemble, for all six cases, and
for all three types of dynamics.  At $t=0.4$ the particle dynamics without
memory indicates a settling with $W$ for all the cases, without any further
change, while the results following from the Maxey-Riley equation predict a
settling with about $W/2$, with a difference between the cases of different
Stokes numbers, and a monotonous increase for $t>0.4$.  On this scale no
difference can be seen between the cases with different excess densities
$\Delta \rho$.
It is interesting to compare the numerical data with the analytic expression
presented for the free fall in still fluids in Section~\ref{sec:still}. To this
end, we have to rescale equations (\ref{eq:free}) and (\ref{eq:solution_appi})
according to the units used for the turbulent flow. Since the time unit in still
fluid can only be $\tau_p$, but in turbulence it is chosen to be the Kolmogorov
time $\tau_{\eta}$, and the Stokes number is exactly $\tau_p/\tau_{\eta}$ (see
Eq.(\ref{St})), the dimensionless time $t'$ of those equation should be
transformed into a $t/St$, where $t$ is the dimensionless time used in all our
equations.  Simultaneously, $v'_z$ of the still fluid case should be replaced by
$W v_z$ in order to be converted to our units.  The different curves in
Fig.~\ref{fig:vel_short} represent the still fluid results (\ref{eq:free}) and
(\ref{eq:solution_appi}) in these units. Since our six parameter sets are
grouped around two Stokes numbers, with which time is scaled, each type of
solution appears with two curves.  A surprising observation is that all points
representing the ensemble averages (symbols) of the turbulent results fall
exactly on the still-fluid curves.

\begin{figure*}[t]
\centering
\includegraphics[scale=.7]{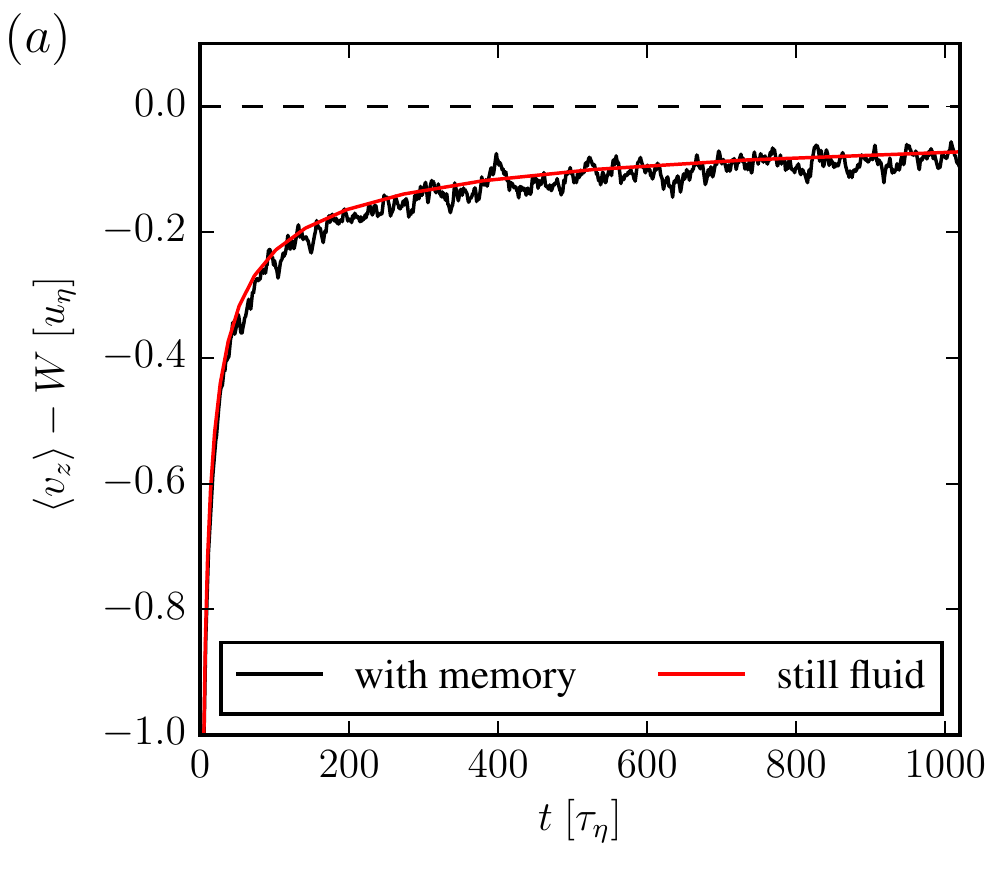}
\includegraphics[scale=.7]{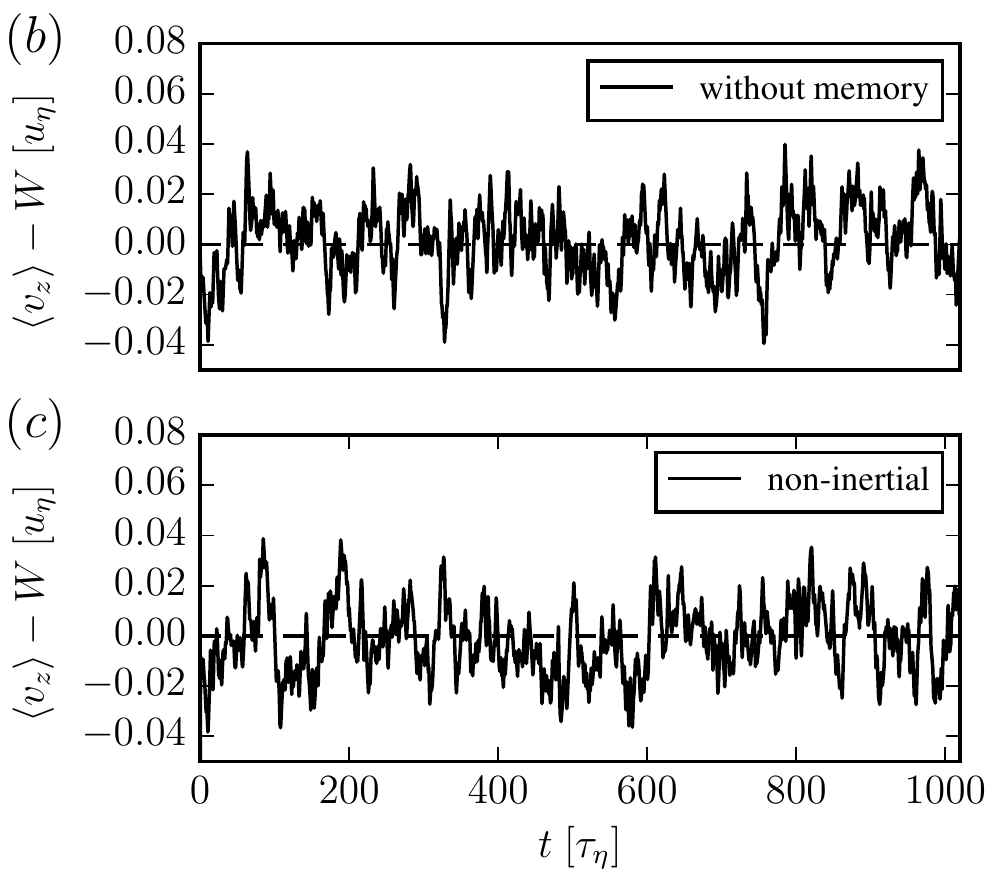}\\

\includegraphics[scale=.7]{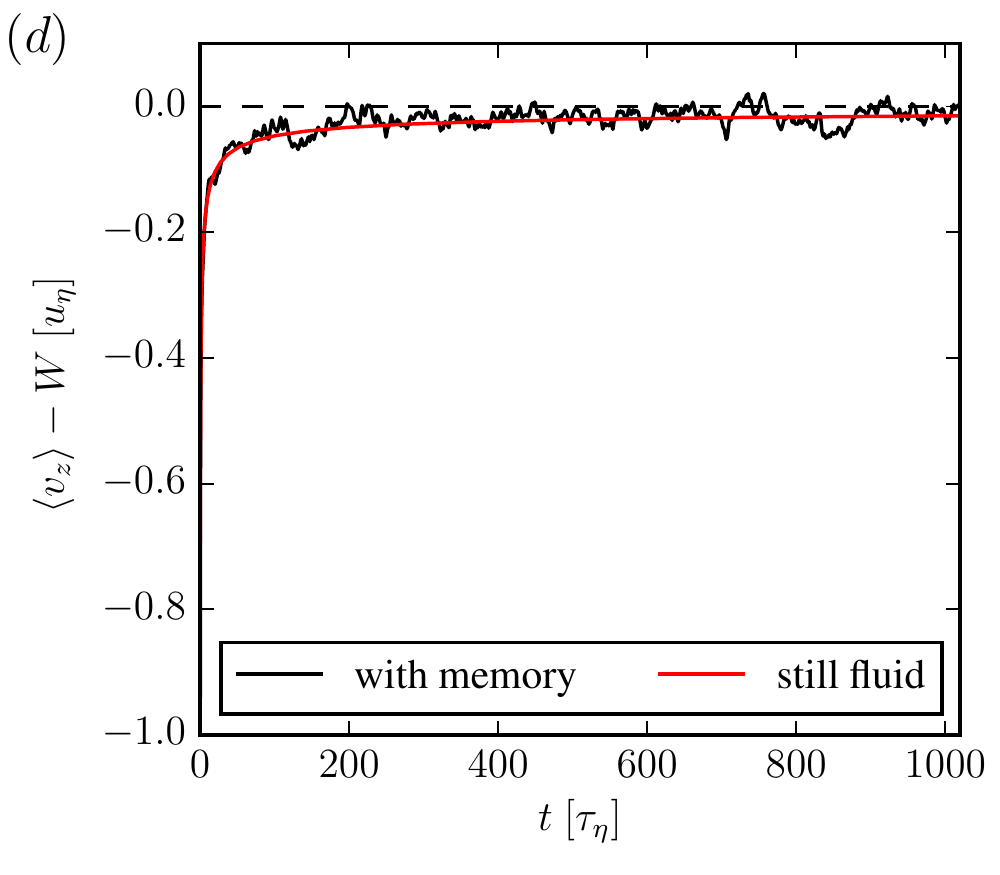}
\includegraphics[scale=.7]{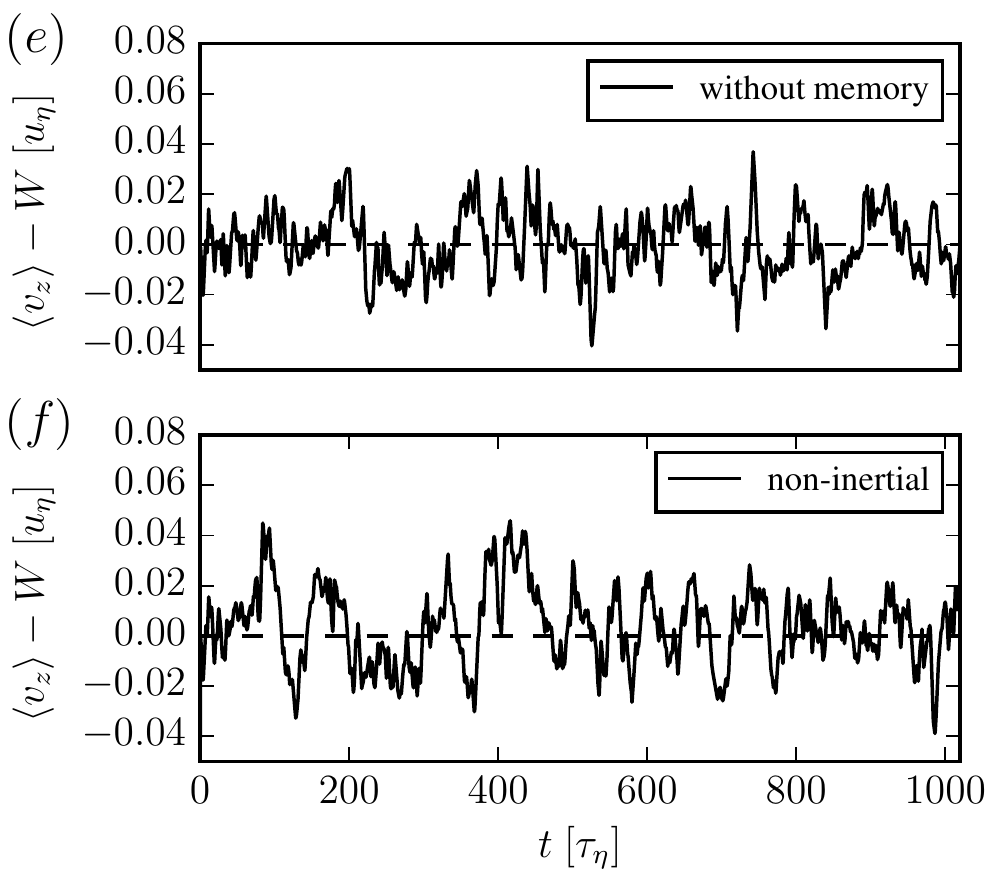}
\caption{Long-term behavior of the velocity difference
  $(\left<v_z\right>-W)$ for cases I ($a, b, c$) and III ($d, e, f$).
  The left column ($a, d$) shows the results of the Maxey-Riley
  equation, the right column those without memory effects. Continuous
  red lines represent (\ref{eq:solution_appa}) an approximate form of
  settling in still fluid, and fits nevertheless very well to the
  turbulent data. }\label{fig:vel_long}
\end{figure*}

To see the long-term behavior, and check if the relation with the still fluid
results hold also on this time scale, we show in Fig.~\ref{fig:vel_long} the
difference between $W$ and the ensemble averaged vertical velocity up to $1020$
Kolmogorov times. Since the large eddy turnover time $T$ in
Table~\ref{tab:sim-para} is about $30 \tau_{\eta}$, $t=1020$ corresponds to
about $34$ turnover times, quite a considerable time span in turbulence. The
results of two cases (I and III) are shown in Fig. 8a,d for the Maxey-Riley
dynamics with memory, in Fig. 8b,e for the dynamics without memory and in
Fig. 8c,f for the non-inertial dynamics. With memory, the ensemble averaged
settling velocity is always below $W$, and has not yet reached a steady value by
the end of the investigated time interval. This is so even for averages taken
over finite time windows of, say, one large eddy turnover time. Such smoothed
time series (not shown) are, however, remarkably close to the result valid in
still fluid. In order to check if this property is not a consequence of the
relatively large settling velocities, we carried out additional simulations with
$10$ times smaller $W$-s but the same Stokes numbers as in Table I. The results
in turbulent flow are found to correlate with the still fluid settling just as
in Figs.~\ref{fig:vel_short},~\ref{fig:vel_long}. In fact, the red lines
represent the function:
\begin{equation}
\left<v_z\right>(t) = W\left[1 - \sqrt{\frac{3 St}{\pi t}}\right],
\label{eq:solution_appa}
\end{equation}
which follows from (\ref{eq:solution_app}) to be valid asymptotically, and the
deviation of $\beta$ from unity can be neglected since all our access densities
are rather small.  This indicates that the decay towards the asymptotic settling
velocity is of power-law type, decaying as one over the square-root of time.
Since such functions are scale-free, no characteristic time can be associated
with them (in contrast e.g. to exponential decays).

Regarding the right column of panels, note the very small scale on the vertical
axes. In all cases the average is zero, meaning that the average settling
velocity over the investigated time interval is $W$, as in still water. The
graphs of the non-inertial and memoryless dynamics are somewhat different,
but they basically represent a random process around zero.  Fluctuations in all
panels are on the order of $0.05$.  These features also hold for the results of
the other four cases not shown here.

\begin{figure}[h]
\centering
\includegraphics[scale=.8]{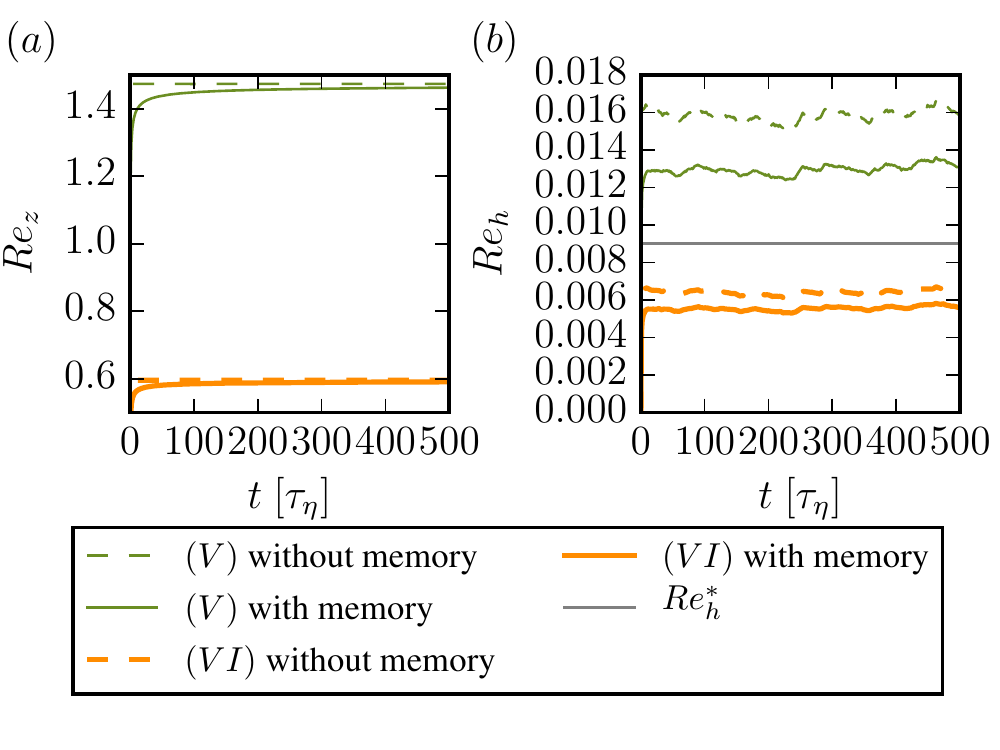}
\caption{Vertical (left panel) and horizontal (right panel) Reynolds numbers as
  a function of time for case V (upper curves) and VI (lower curves). The grey
  horizontal line represents the order of magnitude estimate $Re_h^*$ of
  (\ref{Re*}).}\label{fig:Re_p}
\end{figure}

We also evaluate the slip velocities of the particle ensemble as time series
with and without memory and, based on these define an instantaneous vertical and
horizontal Reynolds number $Re_z(t)$ and $Re_h(t)$ in an analogous way as
$Re_z^*$ and $Re_h^*$ are defined in (\ref{Re*}), but this time with the average
of the modulus of the instantaneous slip velocity
$\left<|\vec{v}(t) -\vec{u}(t)|\right>$. The results obtained for cases V and VI
are summarized in Fig.~\ref{fig:Re_p}.  With memory, the vertical Reynolds
number converges according to a power law to a long-term limit, which is close
to $Re_z^*=W a/ \eta$, a value the other dynamics reach practically
immediately. The order of magnitude of the limiting Reynolds number is unity in
all the cases (the values coincide with those given in Table I).  The horizontal
Reynolds numbers are much smaller than unity. They depend on the access density,
and differ a little bit with and without memory. In any case they happen to be
close to the estimated value $Re_h^*=St \; a/\eta$.  The difference is changing
with W, and we can discover a simple relation
\begin{equation}
\frac{\overline{Re_{hw}} - \overline{Re_{hm}}}{W} = const.
\end{equation}
to hold, where index $m$ and $w$ stand for memory and without memory,
respectively. The value of the constant is found to be about $0.005$ and
$0.0006$ for $St=0.083$ and $St=0.03$, respectively.  It reflects that for
$W \rightarrow 0$ the particles have smaller and smaller excess densities, and
their dynamics approaches that of ideal fluid elements with the zero initial
slip velocity condition used in this paper.

\begin{figure*}[th!]
\centering
\includegraphics[scale=.9]{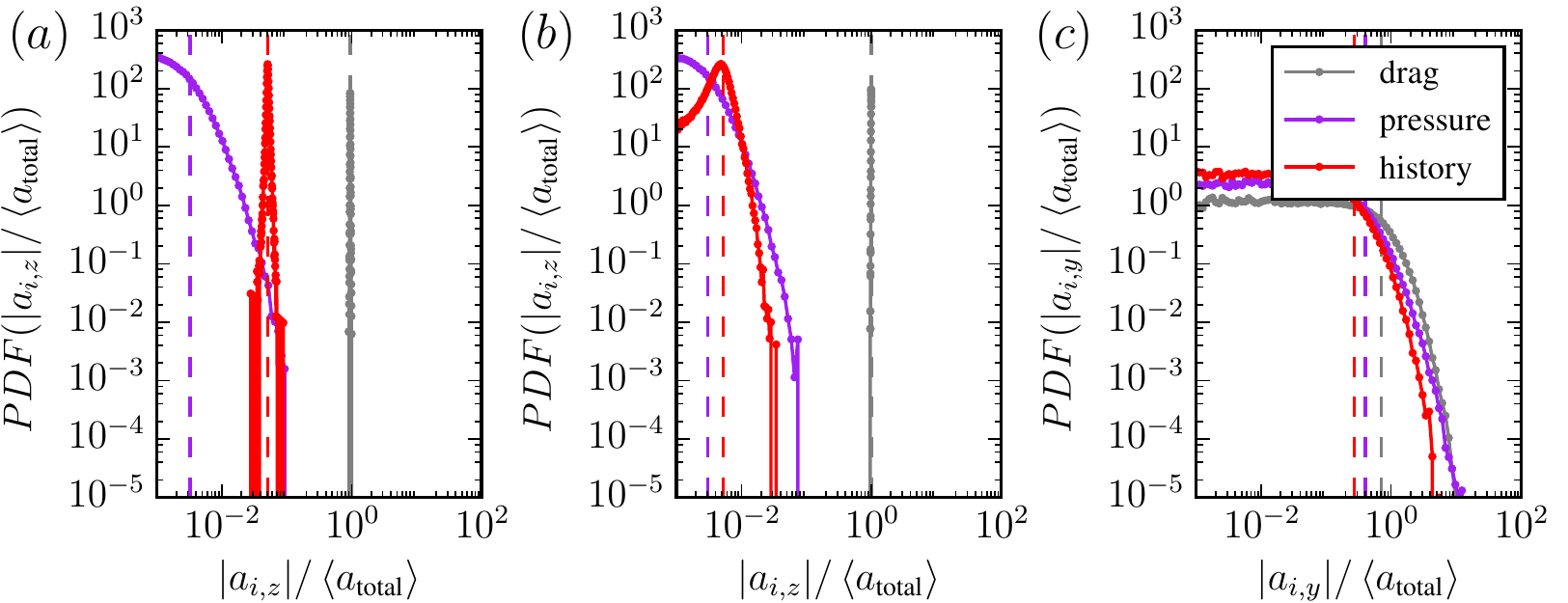}
\caption{Pdf-s of the accelerations $\vec{a}_i$ due to different forces
  (i=pressure, drag or memory) for case IV (a) $z$ component at time
  $t=10.5\; \tau_{\eta}$, (b) $z$ component at $t=1020\; \tau_{\eta}$, (c) $y$
  component at $t=1020\; \tau_{\eta}$. Vertical dashed lines indicate the
  ensemble averages.  $\left<a_{total}\right>$ denotes the ensemble average of
  the modulus of the resultant acceleration.}  \label{fig:acceleration}
\end{figure*}

It is worth also considering the distributions (pdf-s) of the different types of
accelerations.  In Fig.~\ref{fig:acceleration} the acceleration due to the drag,
pressure and history force are plotted, for case IV, at time instants
$t=10.5 \; \tau_{\eta}$ and $t=1020 \; \tau_{\eta}$ for the vertical, and only for
the last instant for the horizontal components.  In the vertical, the drag
dominates, and has a rather narrow distribution. This is due to the fact that
the slip velocity becomes quickly to be of order $W$. On the other hand, the pdf
of the acceleration from the pressure term is rather broad and hardly changes
with time after $t=10.5\; \tau_{\eta}$.  These two pdf-s are found nearly
identical with those in the memoryless equations. Only the pdf of the history
force (red) changes with time rather dramatically: at $t=10.5\; \tau_{\eta}$ it
is sharp an has a much larger average than the pressure contribution. By the end
of the observational period, however, the pdf broadens and becomes shifted
towards smaller values. Its average remains only slightly larger than that of
the pressure.  In the horizontal, the distributions are similar, do not change
too much in time, the averages are ordered as drag, pressure and history with
not very much differences. These pdf-s are rather different from those obtained
without gravity in~\cite{daitche_role_2015}: all distributions are broad there,
the pressure contribution is the largest, those of drag and history are
comparable, and the averages are not separated by several orders of
magnitudes. The closeness of the averages resembles Fig. 10c.

\begin{figure}[h]
\centering
\includegraphics[scale=.6]{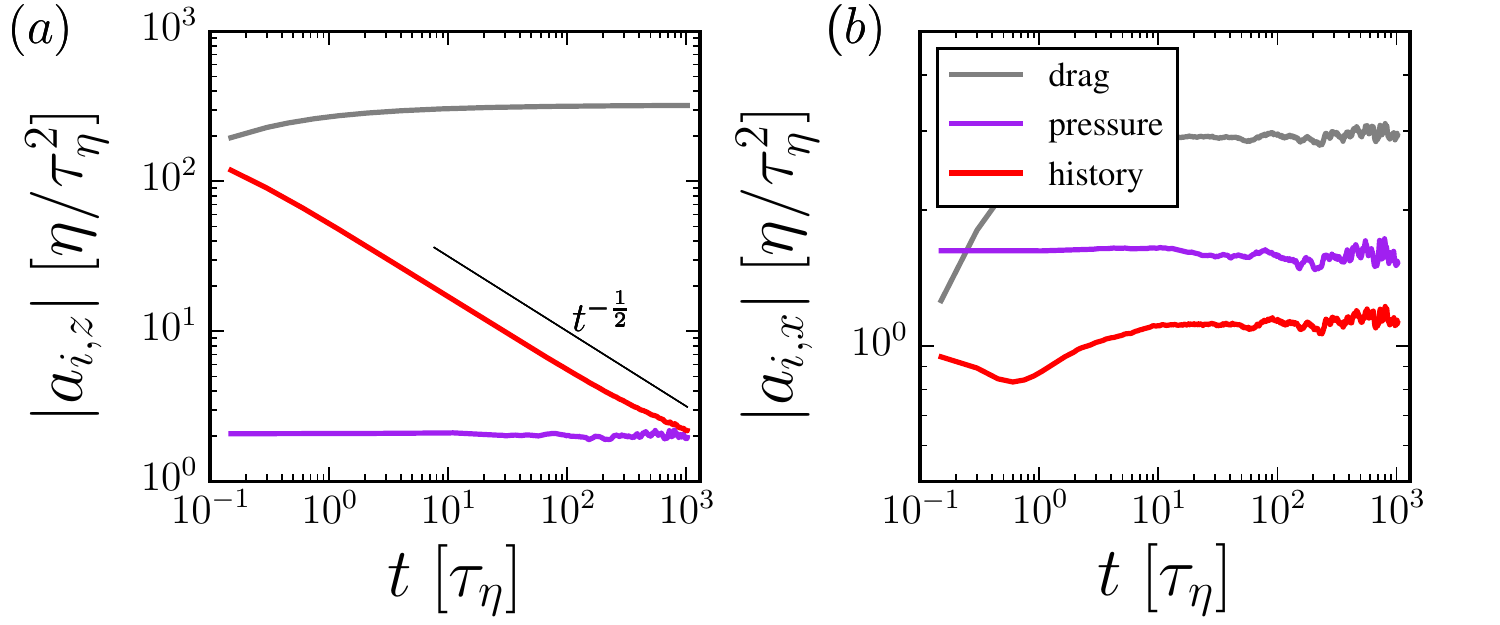}
\caption{Time dependence of the mean of the different pdf-s shown in the
  previous figure. Left panel: vertical, right panel: horizontal components. The
  thin black line in the left panel is of slope $-1/2$ to guide the
  eye. }\label{fig:accelerationt}
\end{figure}

To gain insight into the full time dependence, we plot in
Fig.~\ref{fig:accelerationt} the ensemble average of the pdf-s shown above.  A
striking feature in the vertical components (left panel) is the monotonous decay
of the history force.  Sooner or later, the average of the history force is
likely to become smaller than that of the pressure force.  This internal
degradation of the history force seems to be specific to the sedimentation
dynamics with all the parameters investigated.  Horizontally (right panel),
however, everything is stationary after $t=10.5$. The relatively small values of
the history acceleration explain why no sign of a slow convergence is seen in
Fig.~\ref{fig:Re_p}b. The stationarity of the average pressure acceleration
indicates the stationarity of our turbulent flow. Its order of magnitude is
indeed $u_{\eta}/\tau_{\eta} = \eta/\tau_{\eta}^2$. It is worth noting that the
averages of the accelerations themselves without taking the modulus would all be
zero with the exception to the vertical drag and vertical history acceleration.

\section{Estimating the relevance of the Fax\'en corrections}\label{sec:Faxen}

The Fax\'en corrections are corrections to (\ref{eq:MR}) due to the finite size
of the particle and to the curvature of the flow. They appear as terms
proportional to $a^2 \Delta \vec{u}$ which are added to the slip velocity in the
Stokes drag and in the nominator of the memory integral, as well as to the fluid
velocity in the added mass
term~\cite{maxey_equation_1983,gatignol_faxen_1983}. They appear with a
coefficient $1/6$ and $1/10$, respectively.  We concentrate here on the
correction to the slip velocity and consider the ratio of the average modulus of
the correction to that of the slip velocity
\begin{equation}
C_j=\frac{a^2}{6} \frac{\left<| \Delta \vec{u}_j |\right>}{\left<| \vec{v}_j -\vec{u}_j |\right>},
\label{Cj}
\end{equation}
where index $j$ stands for the Cartesian components $x,y$ or $z$ in this
correction factor. Because of the anisotropy due to gravity, it is worth
treating the horizontal and vertical components separately. Their difference
becomes clear from a simple estimation. Since the characteristic length and
velocity scale of the turbulent flow are $\eta$ and $u_\eta$, respectively, the
Laplacian in any component can be estimated as $u_\eta/\eta^2$.  The slip
velocity in the vertical is approximately $W u_\eta$, while that in the
horizontal is $St u_\eta$, as used in (\ref{Re*h}).  We thus find the estimates
for the vertical and horizontal correction factors
$$
C^*_z=\frac{1}{6} \left( \frac{a}{\eta} \right)^2 \frac{1}{W}, \;\;\;
C^*_h=\frac{1}{6} \left( \frac{a}{\eta} \right)^2 \frac{1}{St}.
$$
Since $W$ is larger than unity in our cases, but $St<1$ (see Table I.), the
relative importance of the Fax\'en corrections is expected to be much smaller in
vertical than in horizontal direction. For our largest particles $a/\eta=1/2$,
and $St=0.083$, thus the estimate $C^*_h$ amounts to a value $0.5$.

\begin{figure}[h]
	\centering
	\includegraphics[scale=.7]{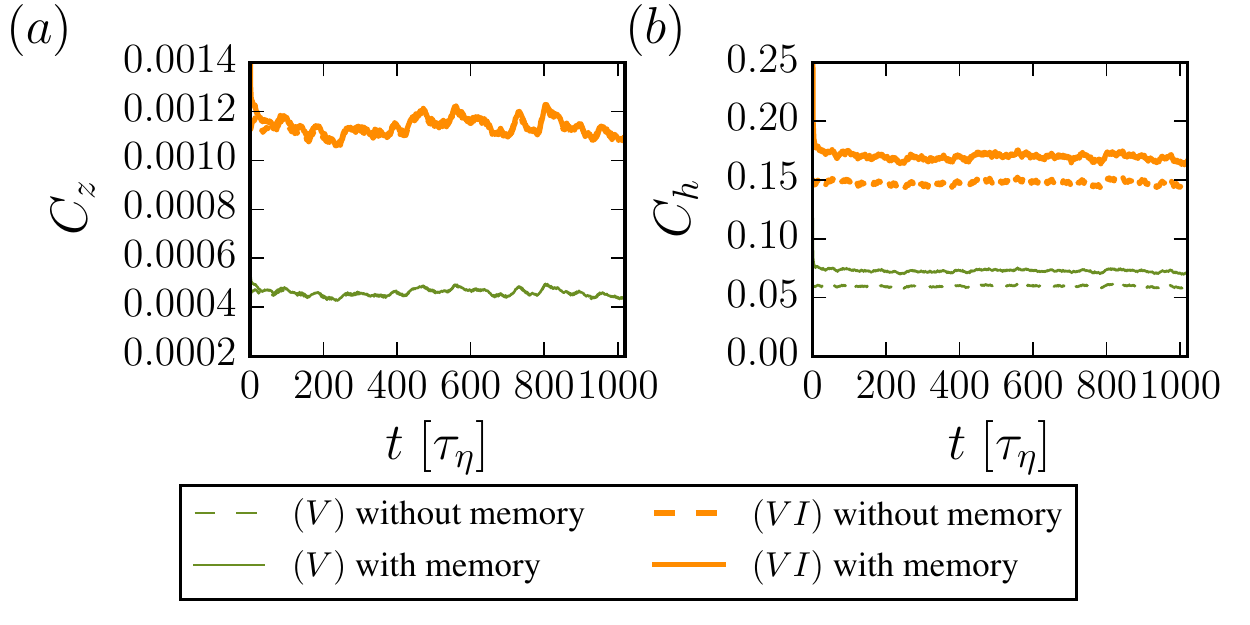}
	\caption{Time dependent correction factor $C_z$ (a) and $C_x$ (b) for
          case (V) and (VI) with memory and without memory.  }\label{fig:Fax}
\end{figure}

We numerically determine the correction factors $C_j$ as functions of time. The
results in the presence of memory are shown in Fig.~\ref{fig:Fax} for $C_z$ and
$C_x$ for cases (V) and (VI) with and without memory. The vertical corrections
(Fig.~\ref{fig:Fax}a) appear to be at most $0.1$ \% consistently, with hardly
any difference with and without memory.  For smaller excess density (case (VI))
the correction is larger. The measured values are about a factor 5 smaller than
the estimates $C^*_z$. In the horizontal, the corrections factor reaches nearly
$20$ \%, but is about a factor 3 smaller than what the estimate $C^*_h$
predicts. There is a measurable difference in the correction with memory and
without, and the former one is consistently larger by about $10$ \%.  The effect
for lighter particles is here stronger again. A comparison with
Fig.~\ref{fig:Re_p} reveals that the tendencies in the Reynolds numbers and in
the correction factors are roughly the opposites.

We thus find that in the vertical Fax\'en corrections can safely be
neglected. In the horizontal, the Fax\'en corrections might be on the same
order, but yet smaller, as the slip velocity.  We found, however, the horizontal
slip velocity to be small compared to unity (see
Fig.~\ref{fig:Re_p}b). Modifying this difference by a factor smaller than unity
does not change the basic observation of the paper that the horizontal Reynolds
number is small, i.e. that the particles follow in the horizontal direction the
fluid motion very closely.

\section{Conclusions}\label{sec:conclusion}

\begin{figure*}[t]
	\centering
	\includegraphics[scale=0.8]{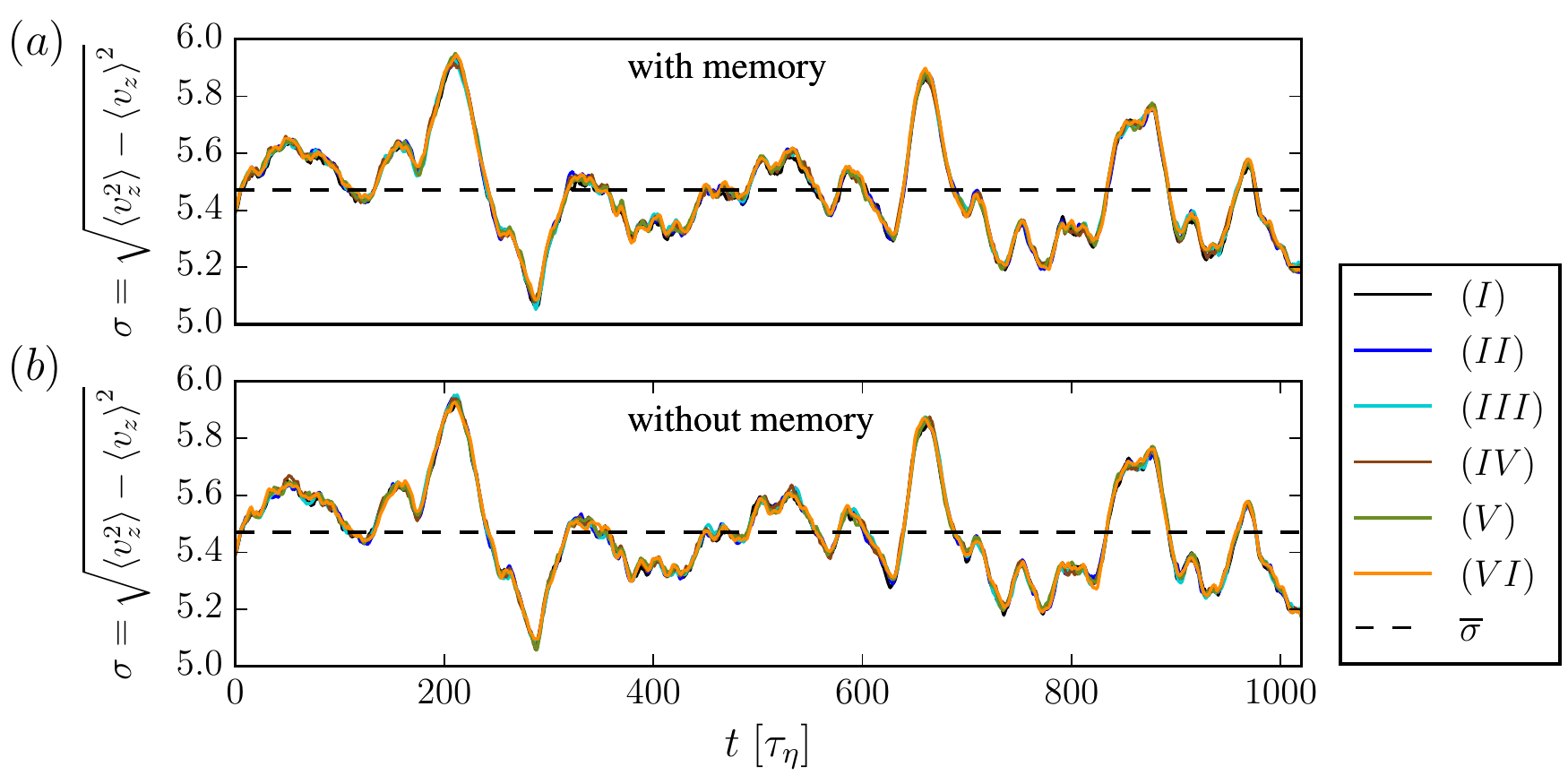}
	\caption{Time dependence of the vertical velocity variance of the
          ensembles, of the different cases marked with different colors. (a):
          with memory, (b) without memory. The horizontal dashed lines mark the
          temporal average $\overline{\sigma}$ of these quantities averaged over
          all cases. These values happen to be identical for the two dynamics,
          and close to the velocity variance of the turbulent
          flow.}\label{fig:vel}
\end{figure*}

Inspired by the sedimentation of marine snow particles in the ocean, we have
studied the impact of the history force on the sedimentation of almost neutrally
buoyant spherical particles in a three-dimensional turbulent flow. Our study is
based on the Maxey-Riley equation and we compared our results with the dynamics
of particles when neglecting the history force, as well as, with non-inertial
particles. We have analyzed 6 parameter sets for marine snow particles
corresponding to typical situations in estuaries and the open ocean. We have
shown that the history force, which introduces a memory, cannot be
neglected.While it leads to large deviations of the trajectories of individual
particles from the ones without memory or of non-inertial particles, the
differences in the horizontal dynamics and spatial extensions of ensemble of
particles are not that large. The most striking effect concerns the vertical
dynamics: when the history force is taken into account, the vertical velocity of
the center of mass of the cloud approaches very slowly a constant settling
velocity, according to a one-over-square-root of time law.

Furthermore, our results indicate that for all three approaches the settling of
small particles, possessing a density not much larger than the one of the fluid,
is surprisingly well described in turbulence by the settling in a still
fluid. The history force leads to a much slower convergence to the settling
velocity, and the limit has not even be reached after more than 1000 Kolmogorov
times. This convergence is of power-law type and we demonstrated a simple,
general expression in the form of (\ref{eq:solution_appa}) to hold.

By contrast, the "ad hoc" dynamics obtained by neglecting the memory converge to
the settling velocity even within one Kolmogorov time unit.  The turbulent
motion of the fluid manifests itself in both cases only in fluctuations around
the settling velocity, which are determined by the properties of the flow. We
illustrate this finding by showing in Fig.~\ref{fig:vel} the time dependence of
the vertical velocity variance for all the cases in a single panel, marked with
different colors. On the scale of $1000$ Kolmogorov units they hardly differ,
and the overall shape is very similar both in the Maxey-Riley equation with
memory (Fig.~\ref{fig:vel}a), and for inertial particles without memory
(Fig.~\ref{fig:vel}b). One hardly sees any difference with naked eye, and this
also holds for the result obtained with non-inertial particles (not
shown). Given that the velocity variance is 5.4 Kolmogorov units in the
turbulent flow (see Table II.), we conclude that it is mainly the flow that
determines the particle dynamics, as the particles are very close in their
densities to that of the fluid. The variance due to particle properties and
advection dynamics only appears in the width of the plotted curves. This width
is about 1\% on this scale, which is in line with the observed velocity
fluctuations being less than 0.05 in Fig.~\ref{fig:vel_long}. Changes in the
variance as well as in the vertical velocity difference happen on a timescale
which is comparable to the large-eddy turnover time of about 30 Kolmogorov
times.

There has been an extensive study of settling of inertial particles in turbulent
fluids by Wang \& Maxey~\cite{wang_settling_1993} although without memory
effects. Though their analysis differs from ours in several aspects, we find it
interesting to compare our results with theirs. Wang \&
Maxey~\cite{wang_settling_1993} have considered only the Stokes drag and the
gravity as the forces acting on their particles, i.e. their study applies to
very heavy particles ($\rho_p >> \rho_f$) instead of the light particles on
which we focus. One of their main achievements consists in the finding that
turbulent motion leads to an additional acceleration of the particles resulting
in an enhanced settling velocity. Their explanation is based on the strongly
inhomogeneous distribution of their heavy particles due to the formation of
preferential concentrations in the flow. Due to inertia the heavy particles are
expelled from the vortices in the flow and whenever they encounter a vortex
during settling they will be accelerated in the direction of its rotation which
moves it into the direction of the downwards motion of the fluid. In our case of
very light particles we {\em do not observe} preferential concentration, the
particles are almost homogeneously distributed. These light particles,
experiencing additionally the pressure and the history force, exhibit a dynamics
which is closer to that of non-inertial particles, for which such a net effect
on the average settling velocity can not be expected. The latter conjecture has
already been formulated by Wang \& Maxey~\cite{wang_settling_1993,
  maxey_gravitational_1987} and our study seems to confirm that. To be able to
observe preferential concentration for such light particles we would have to go
beyond the scope of the Maxey-Riley equation.

Introducing gravity into the dynamics of inertial particles reveals that the
settling velocity appears as another important parameter besides the Stokes
number. One could argue that the effect of the turbulent fluid flow on the
settling of particles could be more pronounced when the settling velocity is
larger than the one for our light particles. However, looking at the particle
Reynolds numbers it turned out that one can distinguish between a vertical
particle Reynolds number $Re_z$ and a horizontal one $Re_h$. Because the
horizontal one scales with the Stokes number which is very small for our cases,
the particle Reynolds number is largely determined by the vertical one which
scales with the settling velocity. These estimates for the two components of the
particle Reynolds number reveal the difficulty in studying particles with larger
settling velocities due to larger densities: the vertical particle Reynolds
number would increase in such a way, that the Maxey-Riley equation would not be
valid anymore. This equation is known to be valid~\cite{daitche_role_2015}
under the assumption that the particle Reynolds number is smaller or
approximately unity, which would be violated for heavy particles.

We find striking differences between the horizontal and the vertical components
of the forces acting on the particle. While the horizontal components of the
drag, the pressure and the history force are almost constant after some
transient time, this applies only to the vertical components of drag and
pressure. The vertical component of the history force, however, becomes smaller
and smaller as time goes by. This can be interpreted as an indicator for that
this force does not have an essential influence on the \emph{asymptotic}
settling velocity.

Let us add a remark on cases when marine aggregates of different sizes and of
different excess densities are considered simultaneously, as a superensemble,
with some size and density distribution.  To understand their typical settling
dynamics, it is worth rewriting the leading term in (\ref{eq:solution_app}) in
dimensional units. After averaging, this leads to

$$
\bigg \langle \frac{W_{\text{settling}}-\left<v_z\right>(t)}{W_{\text{settling}}}
\bigg \rangle_s \approx \frac{\left<a\right>_s}{\sqrt{\pi \nu t}}.
$$

In the turbulent context, $\left<v_z\right>$ means the average vertical velocity
of the particle cloud of a given size and density, and $\left<\right>_s$ stands
for the average taken over the superensemble of different marine
aggregates. Since the right hand side is \emph{ independent} of the
density~\footnote{The excess density is important, of course, for
  $W_{\text{settling}}$, see also (\ref{W}), but not for the ratio investigated here.}
(and also of the fluid properties), the average of the relative deviation from
the asymptotic settling velocity will be proportional to the average size in the
superensemble. The larger this size, the slower the convergence. For an average
size of $1$ mm, and with the viscosity of water, $\nu= 10^{-6}$ in SI units, for
example, the deviation remains more than one percent, for $t<10^4/3$ s, i.e. for
practically one hour. The convergence to a uniform settling velocity is thus
expected to be rather slow also in a superensemble, due to the history force.

Finally we would like to briefly turn to the settling of plankton. These can be
considered as particles of more or less the same excess density as marine snow,
but a factor of 10 smaller in size, with a typical radius of 10 micrometers.
The effect of the history on the settling of plankton was numerically studied by
Olivieri in his thesis~\cite{olivieri_analysis_2013}. He chose two parameter
sets, both with very small $St$ and $W$. The weak effect of gravity leads to
dynamics where the action of all forces is almost isotropic, and the difference
between the horizontal and the vertical directions is small.  Although he
observes some deviations from $W$ for the vertical velocity, he attributes them
to statistical fluctuations and concludes that these small microorganisms will
be carried by the flow as non-inertial tracers.  This is in harmony with our
findings since a change to a=10 micrometers (corresponding to a typical plankton
cell) a factor 30-50 smaller than our aggregates, would make even the
  one-over-square-root type decay to appear very fast.

\section{Acknowledgment}
T.T. acknowledges the support of OTKA grant NK100296 and of the Alexander von
Humboldt Foundation.  We are grateful to George Jackson for illuminating
discussions.  U.F. would like to thank T. T\'el and his group for hospitality
during the stay at E\"otv\"os University Budapest and the Hungarian Academy of
Sciences for financial support.

\appendix

\section{}

\begin{figure*}[t]
\centering
\includegraphics[scale=0.75]{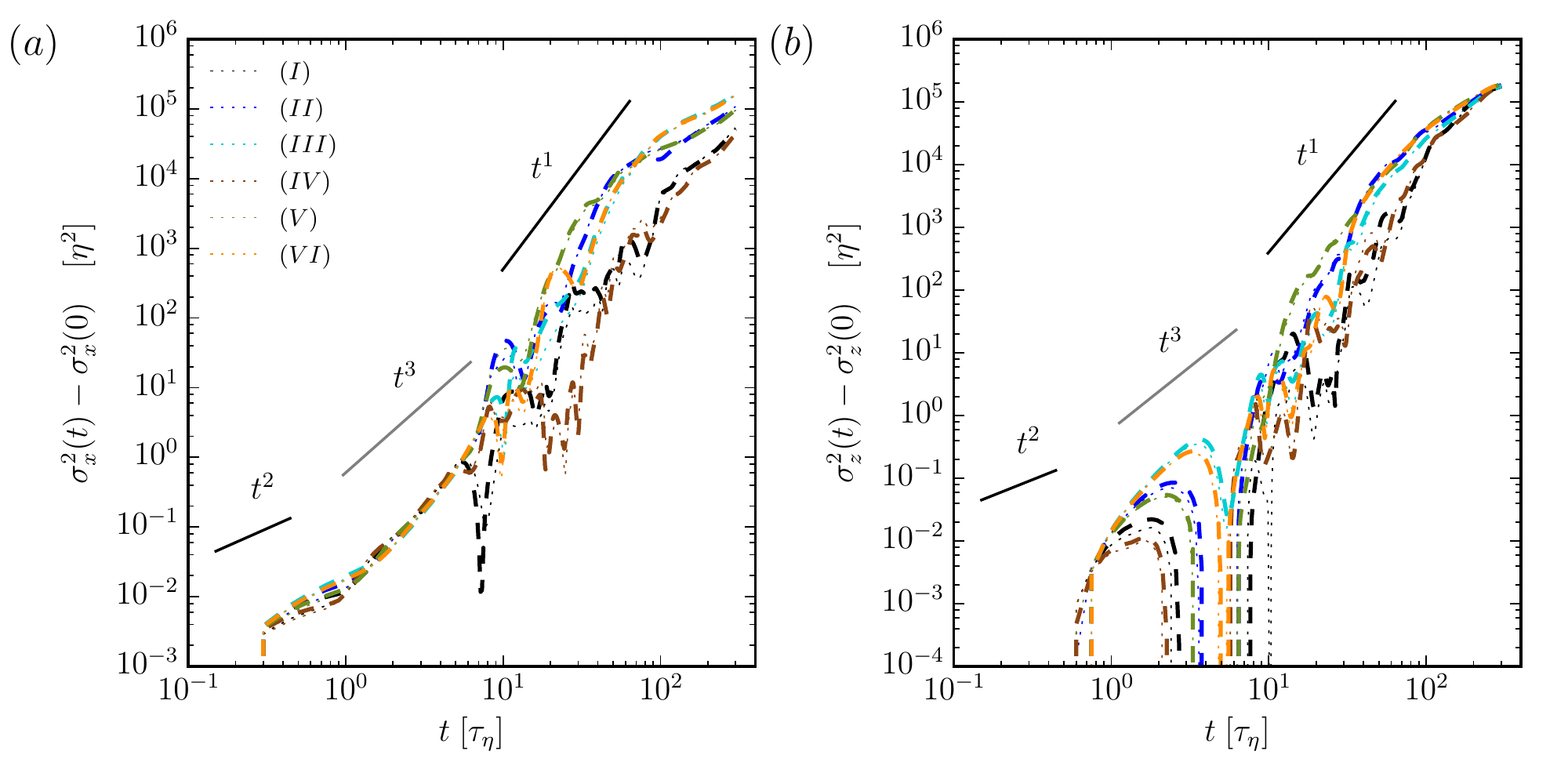}

\caption{ Time-dependence of the variances in the horizontal (x, (a)) and
  vertical (z, (b)) directions in the different cases (coloring and line types
  as in the previous figures) on log-log scales. The ensemble is initiated in a
  box of $L=1 \; \eta$. For clarity, the initial variance
  ($\sigma_x(0) = 1/\sqrt{12} = 0.287\; \eta$) is subtracted.  Straight lines with
  slopes 1,2 and 3 are overlaid to lead the eye. }\label{fig:sigma_appendix}
\end{figure*}

In order to test how an initially strongly localized ensemble behaves, we
carried out a single extra simulation with $N_p=5 \cdot 10^4$ particles
uniformly distributed at time $t=0$ in a box of size $L=1 \cdot \eta$ centered
at the origin, with zero initial slip velocities. Since the differences among
the three types of dynamics are minor for the ensemble variances, here we chose
non-inertial and inertial particles without memory only, since they require
lower computational demand. The simulation is carried out up to 300 $\tau_\eta$.
The variances are shown in Fig.~\ref{fig:sigma_appendix}. In contrast to
Fig.~\ref{fig:sigma}, here a clear intermediate time scaling with $t^3$ can be
found, i.e., Richardson's scaling becomes observable. A difference between the
horizontal and the vertical dynamics is that in the latter (panel b) ballistic
behaviour is not observable in the data.

It is worth determining the dimensionless crossover time $t_0/\tau_\eta$ for
this case, too. Along the lines applied in Section V, we find
\begin{equation}
\frac{t_0}{\tau_\eta} = (3 \sigma^2_x(0))^{1/3}=(3/12)^{1/3}= 0.630.
\end{equation}
This corresponds precisely to the time where a crossover from the quadratic to
the cubic, Richardson scaling takes place. Note that the second crossover from
the Richardson to a diffusive behavior occurs at about 30 $\tau_\eta$, i.e. at
the eddy turnover time $T$. The lack of the Richardson regime in Fig. 6 is due
to the fact that the turnover time $t_0$ is larger than $T$, and hence there is
no "space" for the cubic behavior.

\bibliographystyle{apsrev4-1}
\bibliography{HF-marinesnow.bib}

\begin{thebibliography}{61}%
\makeatletter
\providecommand \@ifxundefined [1]{%
 \@ifx{#1\undefined}
}%
\providecommand \@ifnum [1]{%
 \ifnum #1\expandafter \@firstoftwo
 \else \expandafter \@secondoftwo
 \fi
}%
\providecommand \@ifx [1]{%
 \ifx #1\expandafter \@firstoftwo
 \else \expandafter \@secondoftwo
 \fi
}%
\providecommand \natexlab [1]{#1}%
\providecommand \enquote  [1]{``#1''}%
\providecommand \bibnamefont  [1]{#1}%
\providecommand \bibfnamefont [1]{#1}%
\providecommand \citenamefont [1]{#1}%
\providecommand \href@noop [0]{\@secondoftwo}%
\providecommand \href [0]{\begingroup \@sanitize@url \@href}%
\providecommand \@href[1]{\@@startlink{#1}\@@href}%
\providecommand \@@href[1]{\endgroup#1\@@endlink}%
\providecommand \@sanitize@url [0]{\catcode `\\12\catcode `\$12\catcode
  `\&12\catcode `\#12\catcode `\^12\catcode `\_12\catcode `\%12\relax}%
\providecommand \@@startlink[1]{}%
\providecommand \@@endlink[0]{}%
\providecommand \url  [0]{\begingroup\@sanitize@url \@url }%
\providecommand \@url [1]{\endgroup\@href {#1}{\urlprefix }}%
\providecommand \urlprefix  [0]{URL }%
\providecommand \Eprint [0]{\href }%
\providecommand \doibase [0]{http://dx.doi.org/}%
\providecommand \selectlanguage [0]{\@gobble}%
\providecommand \bibinfo  [0]{\@secondoftwo}%
\providecommand \bibfield  [0]{\@secondoftwo}%
\providecommand \translation [1]{[#1]}%
\providecommand \BibitemOpen [0]{}%
\providecommand \bibitemStop [0]{}%
\providecommand \bibitemNoStop [0]{.\EOS\space}%
\providecommand \EOS [0]{\spacefactor3000\relax}%
\providecommand \BibitemShut  [1]{\csname bibitem#1\endcsname}%
\let\auto@bib@innerbib\@empty
\bibitem [{\citenamefont {Tanga}\ and\ \citenamefont
  {Provenzale}(1994)}]{tanga_dynamics_1994}%
  \BibitemOpen
  \bibfield  {author} {\bibinfo {author} {\bibfnamefont {P.}~\bibnamefont
  {Tanga}}\ and\ \bibinfo {author} {\bibfnamefont {A.}~\bibnamefont
  {Provenzale}},\ }\href {\doibase 10.1016/0167-2789(94)90259-3} {\bibfield
  {journal} {\bibinfo  {journal} {Physica D: Nonlinear Phenomena}\ }\textbf
  {\bibinfo {volume} {76}},\ \bibinfo {pages} {202} (\bibinfo {year}
  {1994})}\BibitemShut {NoStop}%
\bibitem [{\citenamefont {Belmonte}\ \emph {et~al.}(2001)\citenamefont
  {Belmonte}, \citenamefont {Jacobsen},\ and\ \citenamefont
  {Jayaraman}}]{belmonte_monotone_2001}%
  \BibitemOpen
  \bibfield  {author} {\bibinfo {author} {\bibfnamefont {A.}~\bibnamefont
  {Belmonte}}, \bibinfo {author} {\bibfnamefont {J.}~\bibnamefont {Jacobsen}},
  \ and\ \bibinfo {author} {\bibfnamefont {A.}~\bibnamefont {Jayaraman}},\
  }\href {http://scholarship.claremont.edu/hmc_fac_pub/761} {\bibfield
  {journal} {\bibinfo  {journal} {All HMC Faculty Publications and Research}\ }
  (\bibinfo {year} {2001})}\BibitemShut {NoStop}%
\bibitem [{\citenamefont {Daitche}\ and\ \citenamefont
  {Tél}(2011)}]{daitche_memory_2011}%
  \BibitemOpen
  \bibfield  {author} {\bibinfo {author} {\bibfnamefont {A.}~\bibnamefont
  {Daitche}}\ and\ \bibinfo {author} {\bibfnamefont {T.}~\bibnamefont {Tél}},\
  }\href@noop {} {\bibfield  {journal} {\bibinfo  {journal} {Physical Review
  Letters}\ }\textbf {\bibinfo {volume} {107}},\ \bibinfo {pages} {244501}
  (\bibinfo {year} {2011})},\ \bibinfo {note} {00007}\BibitemShut {NoStop}%
\bibitem [{\citenamefont {Guseva}\ \emph {et~al.}(2013)\citenamefont {Guseva},
  \citenamefont {Feudel},\ and\ \citenamefont {Tél}}]{guseva_influence_2013}%
  \BibitemOpen
  \bibfield  {author} {\bibinfo {author} {\bibfnamefont {K.}~\bibnamefont
  {Guseva}}, \bibinfo {author} {\bibfnamefont {U.}~\bibnamefont {Feudel}}, \
  and\ \bibinfo {author} {\bibfnamefont {T.}~\bibnamefont {Tél}},\ }\href
  {\doibase 10.1103/PhysRevE.88.042909} {\bibfield  {journal} {\bibinfo
  {journal} {Physical Review E}\ }\textbf {\bibinfo {volume} {88}},\ \bibinfo
  {pages} {042909} (\bibinfo {year} {2013})}\BibitemShut {NoStop}%
\bibitem [{\citenamefont {Daitche}\ and\ \citenamefont
  {Tél}(2014)}]{daitche_memory_2014}%
  \BibitemOpen
  \bibfield  {author} {\bibinfo {author} {\bibfnamefont {A.}~\bibnamefont
  {Daitche}}\ and\ \bibinfo {author} {\bibfnamefont {T.}~\bibnamefont {Tél}},\
  }\href {\doibase 10.1088/1367-2630/16/7/073008} {\bibfield  {journal}
  {\bibinfo  {journal} {New Journal of Physics}\ }\textbf {\bibinfo {volume}
  {16}},\ \bibinfo {pages} {073008} (\bibinfo {year} {2014})}\BibitemShut
  {NoStop}%
\bibitem [{\citenamefont {Farazmand}\ and\ \citenamefont
  {Haller}(2015)}]{farazmand_maxeyriley_2015}%
  \BibitemOpen
  \bibfield  {author} {\bibinfo {author} {\bibfnamefont {M.}~\bibnamefont
  {Farazmand}}\ and\ \bibinfo {author} {\bibfnamefont {G.}~\bibnamefont
  {Haller}},\ }\href {\doibase 10.1016/j.nonrwa.2014.08.002} {\bibfield
  {journal} {\bibinfo  {journal} {Nonlinear Analysis: Real World Applications}\
  }\textbf {\bibinfo {volume} {22}},\ \bibinfo {pages} {98} (\bibinfo {year}
  {2015})}\BibitemShut {NoStop}%
\bibitem [{\citenamefont {Mordant}\ and\ \citenamefont
  {Pinton}(2000)}]{mordant_velocity_2000}%
  \BibitemOpen
  \bibfield  {author} {\bibinfo {author} {\bibfnamefont {N.}~\bibnamefont
  {Mordant}}\ and\ \bibinfo {author} {\bibfnamefont {J.-F.}\ \bibnamefont
  {Pinton}},\ }\href@noop {} {\bibfield  {journal} {\bibinfo  {journal} {The
  European Physical Journal B}\ }\textbf {\bibinfo {volume} {18}},\ \bibinfo
  {pages} {343} (\bibinfo {year} {2000})}\BibitemShut {NoStop}%
\bibitem [{\citenamefont {Candelier}\ \emph {et~al.}(2004)\citenamefont
  {Candelier}, \citenamefont {Angilella},\ and\ \citenamefont
  {Souhar}}]{candelier_effect_2004}%
  \BibitemOpen
  \bibfield  {author} {\bibinfo {author} {\bibfnamefont {F.}~\bibnamefont
  {Candelier}}, \bibinfo {author} {\bibfnamefont {J.~R.}\ \bibnamefont
  {Angilella}}, \ and\ \bibinfo {author} {\bibfnamefont {M.}~\bibnamefont
  {Souhar}},\ }\href@noop {} {\bibfield  {journal} {\bibinfo  {journal}
  {Physics of Fluids}\ }\textbf {\bibinfo {volume} {16}},\ \bibinfo {pages}
  {1765} (\bibinfo {year} {2004})}\BibitemShut {NoStop}%
\bibitem [{\citenamefont {Vojir}\ and\ \citenamefont
  {Michaelides}(1994)}]{vojir_effect_1994}%
  \BibitemOpen
  \bibfield  {author} {\bibinfo {author} {\bibfnamefont {D.}~\bibnamefont
  {Vojir}}\ and\ \bibinfo {author} {\bibfnamefont {E.}~\bibnamefont
  {Michaelides}},\ }\href@noop {} {\bibfield  {journal} {\bibinfo  {journal}
  {International Journal of Multiphase Flow}\ }\textbf {\bibinfo {volume}
  {20}},\ \bibinfo {pages} {547} (\bibinfo {year} {1994})},\ \bibinfo {note}
  {00077}\BibitemShut {NoStop}%
\bibitem [{\citenamefont {Langlois}\ \emph {et~al.}(2015)\citenamefont
  {Langlois}, \citenamefont {Farazmand},\ and\ \citenamefont
  {Haller}}]{langlois_asymptotic_2015}%
  \BibitemOpen
  \bibfield  {author} {\bibinfo {author} {\bibfnamefont {G.~P.}\ \bibnamefont
  {Langlois}}, \bibinfo {author} {\bibfnamefont {M.}~\bibnamefont {Farazmand}},
  \ and\ \bibinfo {author} {\bibfnamefont {G.}~\bibnamefont {Haller}},\ }\href
  {\doibase 10.1007/s00332-015-9250-0} {\bibfield  {journal} {\bibinfo
  {journal} {Journal of Nonlinear Science}\ ,\ \bibinfo {pages} {1}} (\bibinfo
  {year} {2015})}\BibitemShut {NoStop}%
\bibitem [{\citenamefont {Daitche}(2015)}]{daitche_role_2015}%
  \BibitemOpen
  \bibfield  {author} {\bibinfo {author} {\bibfnamefont {A.}~\bibnamefont
  {Daitche}},\ }\href {\doibase 10.1017/jfm.2015.551} {\bibfield  {journal}
  {\bibinfo  {journal} {Journal of Fluid Mechanics}\ }\textbf {\bibinfo
  {volume} {782}},\ \bibinfo {pages} {567} (\bibinfo {year}
  {2015})}\BibitemShut {NoStop}%
\bibitem [{\citenamefont {Maxey}\ and\ \citenamefont
  {Riley}(1983)}]{maxey_equation_1983}%
  \BibitemOpen
  \bibfield  {author} {\bibinfo {author} {\bibfnamefont {M.~R.}\ \bibnamefont
  {Maxey}}\ and\ \bibinfo {author} {\bibfnamefont {J.~J.}\ \bibnamefont
  {Riley}},\ }\href@noop {} {\bibfield  {journal} {\bibinfo  {journal} {Physics
  of Fluids}\ }\textbf {\bibinfo {volume} {26}},\ \bibinfo {pages} {883}
  (\bibinfo {year} {1983})}\BibitemShut {NoStop}%
\bibitem [{\citenamefont {Gatignol}(1983)}]{gatignol_faxen_1983}%
  \BibitemOpen
  \bibfield  {author} {\bibinfo {author} {\bibfnamefont {R.}~\bibnamefont
  {Gatignol}},\ }\href
  {https://www.researchgate.net/publication/233845292_The_Faxn_formulae_for_a_rigid_particle_in_an_unsteady_non-uniform_Stokes_flow}
  {\bibfield  {journal} {\bibinfo  {journal} {J. Mec. Theor. Appl}\ }\textbf
  {\bibinfo {volume} {1}},\ \bibinfo {pages} {143} (\bibinfo {year}
  {1983})}\BibitemShut {NoStop}%
\bibitem [{\citenamefont {Auton}\ \emph {et~al.}(1988)\citenamefont {Auton},
  \citenamefont {Hunt},\ and\ \citenamefont {Prud'Homme}}]{auton_force_1988}%
  \BibitemOpen
  \bibfield  {author} {\bibinfo {author} {\bibfnamefont {T.~R.}\ \bibnamefont
  {Auton}}, \bibinfo {author} {\bibfnamefont {J.~C.~R.}\ \bibnamefont {Hunt}},
  \ and\ \bibinfo {author} {\bibfnamefont {M.}~\bibnamefont {Prud'Homme}},\
  }\href@noop {} {\bibfield  {journal} {\bibinfo  {journal} {Journal of Fluid
  Mechanics}\ }\textbf {\bibinfo {volume} {197}},\ \bibinfo {pages} {241}
  (\bibinfo {year} {1988})},\ \bibinfo {note} {00348}\BibitemShut {NoStop}%
\bibitem [{\citenamefont {Basset}(1888)}]{basset_treatise_1888}%
  \BibitemOpen
  \bibfield  {author} {\bibinfo {author} {\bibfnamefont {A.~B.}\ \bibnamefont
  {Basset}},\ }\href@noop {} {\emph {\bibinfo {title} {A {Treatise} on
  {Hydrodynamics}}}}\ (\bibinfo  {publisher} {Deighton, Bell and Co.},\
  \bibinfo {year} {1888})\BibitemShut {NoStop}%
\bibitem [{\citenamefont {van Hinsberg}\ \emph {et~al.}(2011)\citenamefont {van
  Hinsberg}, \citenamefont {ten Thije~Boonkkamp},\ and\ \citenamefont
  {Clercx}}]{van_hinsberg_efficient_2011}%
  \BibitemOpen
  \bibfield  {author} {\bibinfo {author} {\bibfnamefont {M.}~\bibnamefont {van
  Hinsberg}}, \bibinfo {author} {\bibfnamefont {J.}~\bibnamefont {ten
  Thije~Boonkkamp}}, \ and\ \bibinfo {author} {\bibfnamefont {H.}~\bibnamefont
  {Clercx}},\ }\href@noop {} {\bibfield  {journal} {\bibinfo  {journal}
  {Journal of Computational Physics}\ }\textbf {\bibinfo {volume} {230}},\
  \bibinfo {pages} {1465} (\bibinfo {year} {2011})},\ \bibinfo {note}
  {00028}\BibitemShut {NoStop}%
\bibitem [{\citenamefont {Daitche}(2013)}]{daitche_advection_2013}%
  \BibitemOpen
  \bibfield  {author} {\bibinfo {author} {\bibfnamefont {A.}~\bibnamefont
  {Daitche}},\ }\href {\doibase 10.1016/j.jcp.2013.07.024} {\bibfield
  {journal} {\bibinfo  {journal} {Journal of Computational Physics}\ }\textbf
  {\bibinfo {volume} {254}},\ \bibinfo {pages} {93} (\bibinfo {year}
  {2013})}\BibitemShut {NoStop}%
\bibitem [{\citenamefont {Toegel}\ \emph {et~al.}(2006)\citenamefont {Toegel},
  \citenamefont {Luther},\ and\ \citenamefont {Lohse}}]{toegel_viscosity_2006}%
  \BibitemOpen
  \bibfield  {author} {\bibinfo {author} {\bibfnamefont {R.}~\bibnamefont
  {Toegel}}, \bibinfo {author} {\bibfnamefont {S.}~\bibnamefont {Luther}}, \
  and\ \bibinfo {author} {\bibfnamefont {D.}~\bibnamefont {Lohse}},\
  }\href@noop {} {\bibfield  {journal} {\bibinfo  {journal} {Physical Review
  Letters}\ }\textbf {\bibinfo {volume} {96}},\ \bibinfo {pages} {114301}
  (\bibinfo {year} {2006})}\BibitemShut {NoStop}%
\bibitem [{\citenamefont {Garbin}\ \emph {et~al.}(2009)\citenamefont {Garbin},
  \citenamefont {Dollet}, \citenamefont {Overvelde}, \citenamefont {Cojoc},
  \citenamefont {Di~Fabrizio}, \citenamefont {van Wijngaarden}, \citenamefont
  {Prosperetti}, \citenamefont {de~Jong}, \citenamefont {Lohse},\ and\
  \citenamefont {Versluis}}]{garbin_history_2009}%
  \BibitemOpen
  \bibfield  {author} {\bibinfo {author} {\bibfnamefont {V.}~\bibnamefont
  {Garbin}}, \bibinfo {author} {\bibfnamefont {B.}~\bibnamefont {Dollet}},
  \bibinfo {author} {\bibfnamefont {M.}~\bibnamefont {Overvelde}}, \bibinfo
  {author} {\bibfnamefont {D.}~\bibnamefont {Cojoc}}, \bibinfo {author}
  {\bibfnamefont {E.}~\bibnamefont {Di~Fabrizio}}, \bibinfo {author}
  {\bibfnamefont {L.}~\bibnamefont {van Wijngaarden}}, \bibinfo {author}
  {\bibfnamefont {A.}~\bibnamefont {Prosperetti}}, \bibinfo {author}
  {\bibfnamefont {N.}~\bibnamefont {de~Jong}}, \bibinfo {author} {\bibfnamefont
  {D.}~\bibnamefont {Lohse}}, \ and\ \bibinfo {author} {\bibfnamefont
  {M.}~\bibnamefont {Versluis}},\ }\href@noop {} {\bibfield  {journal}
  {\bibinfo  {journal} {Physics of Fluids}\ }\textbf {\bibinfo {volume} {21}},\
  \bibinfo {pages} {092003} (\bibinfo {year} {2009})},\ \bibinfo {note}
  {00015}\BibitemShut {NoStop}%
\bibitem [{\citenamefont {Bergougnoux}\ \emph {et~al.}(2014)\citenamefont
  {Bergougnoux}, \citenamefont {Bouchet}, \citenamefont {Lopez},\ and\
  \citenamefont {Guazzelli}}]{bergougnoux_motion_2014}%
  \BibitemOpen
  \bibfield  {author} {\bibinfo {author} {\bibfnamefont {L.}~\bibnamefont
  {Bergougnoux}}, \bibinfo {author} {\bibfnamefont {G.}~\bibnamefont
  {Bouchet}}, \bibinfo {author} {\bibfnamefont {D.}~\bibnamefont {Lopez}}, \
  and\ \bibinfo {author} {\bibfnamefont {E.}~\bibnamefont {Guazzelli}},\ }\href
  {\doibase 10.1063/1.4895736} {\bibfield  {journal} {\bibinfo  {journal}
  {Physics of Fluids (1994-present)}\ }\textbf {\bibinfo {volume} {26}},\
  \bibinfo {pages} {093302} (\bibinfo {year} {2014})}\BibitemShut {NoStop}%
\bibitem [{\citenamefont {Aartrijk}\ and\ \citenamefont
  {Clercx}(2010)}]{aartrijk_vertical_2010}%
  \BibitemOpen
  \bibfield  {author} {\bibinfo {author} {\bibfnamefont {M.~v.}\ \bibnamefont
  {Aartrijk}}\ and\ \bibinfo {author} {\bibfnamefont {H.~J.~H.}\ \bibnamefont
  {Clercx}},\ }\href {\doibase 10.1063/1.3291678} {\bibfield  {journal}
  {\bibinfo  {journal} {Physics of Fluids (1994-present)}\ }\textbf {\bibinfo
  {volume} {22}},\ \bibinfo {pages} {013301} (\bibinfo {year}
  {2010})}\BibitemShut {NoStop}%
\bibitem [{\citenamefont {Olivieri}(2013)}]{olivieri_analysis_2013}%
  \BibitemOpen
  \bibfield  {author} {\bibinfo {author} {\bibfnamefont {S.}~\bibnamefont
  {Olivieri}},\ }\emph {\bibinfo {title} {Analysis of the {Forces} {Acting} on
  {Particles} in {Homogeneous} {Isotroic} {Turbulence}}},\ \href@noop {} {Ph.D.
  thesis},\ \bibinfo  {school} {Università degli Studi di Genova}, \bibinfo
  {address} {Genova} (\bibinfo {year} {2013})\BibitemShut {NoStop}%
\bibitem [{\citenamefont {Mann}\ and\ \citenamefont
  {Lazier}(2005)}]{mann_dynamics_2005}%
  \BibitemOpen
  \bibfield  {author} {\bibinfo {author} {\bibfnamefont {K.}~\bibnamefont
  {Mann}}\ and\ \bibinfo {author} {\bibfnamefont {J.}~\bibnamefont {Lazier}},\
  }\href@noop {} {\emph {\bibinfo {title} {Dynamics of {Marine} {Ecosystems}:
  {Biological}-{Physical} {Interactions} in the {Oceans}}}},\ \bibinfo
  {edition} {3rd}\ ed.\ (\bibinfo  {publisher} {Wiley-Blackwell},\ \bibinfo
  {year} {2005})\BibitemShut {NoStop}%
\bibitem [{\citenamefont {De~La~Rocha}\ and\ \citenamefont
  {Passow}(2007)}]{de_la_rocha_factors_2007}%
  \BibitemOpen
  \bibfield  {author} {\bibinfo {author} {\bibfnamefont {C.~L.}\ \bibnamefont
  {De~La~Rocha}}\ and\ \bibinfo {author} {\bibfnamefont {U.}~\bibnamefont
  {Passow}},\ }\href@noop {} {\bibfield  {journal} {\bibinfo  {journal} {Deep
  Sea Research Part II: Topical Studies in Oceanography}\ }\textbf {\bibinfo
  {volume} {54}},\ \bibinfo {pages} {639} (\bibinfo {year} {2007})},\ \bibinfo
  {note} {00086}\BibitemShut {NoStop}%
\bibitem [{\citenamefont {Riley}\ \emph {et~al.}(2012)\citenamefont {Riley},
  \citenamefont {Sanders}, \citenamefont {Marsay}, \citenamefont {Le~Moigne},
  \citenamefont {Achterberg},\ and\ \citenamefont
  {Poulton}}]{riley_relative_2012}%
  \BibitemOpen
  \bibfield  {author} {\bibinfo {author} {\bibfnamefont {J.~S.}\ \bibnamefont
  {Riley}}, \bibinfo {author} {\bibfnamefont {R.}~\bibnamefont {Sanders}},
  \bibinfo {author} {\bibfnamefont {C.}~\bibnamefont {Marsay}}, \bibinfo
  {author} {\bibfnamefont {F.~a.~C.}\ \bibnamefont {Le~Moigne}}, \bibinfo
  {author} {\bibfnamefont {E.~P.}\ \bibnamefont {Achterberg}}, \ and\ \bibinfo
  {author} {\bibfnamefont {A.~J.}\ \bibnamefont {Poulton}},\ }\href@noop {}
  {\bibfield  {journal} {\bibinfo  {journal} {Global Biogeochemical Cycles}\
  }\textbf {\bibinfo {volume} {26}} (\bibinfo {year} {2012})},\ \bibinfo {note}
  {00009}\BibitemShut {NoStop}%
\bibitem [{\citenamefont {Stone}(2010)}]{stone_invisible_2010}%
  \BibitemOpen
  \bibfield  {author} {\bibinfo {author} {\bibfnamefont {R.}~\bibnamefont
  {Stone}},\ }\href {\doibase 10.1126/science.328.5985.1476} {\bibfield
  {journal} {\bibinfo  {journal} {Science}\ }\textbf {\bibinfo {volume}
  {328}},\ \bibinfo {pages} {1476} (\bibinfo {year} {2010})}\BibitemShut
  {NoStop}%
\bibitem [{\citenamefont {Passow}\ and\ \citenamefont
  {Carlson}(2012)}]{passow_biological_2012}%
  \BibitemOpen
  \bibfield  {author} {\bibinfo {author} {\bibfnamefont {U.}~\bibnamefont
  {Passow}}\ and\ \bibinfo {author} {\bibfnamefont {C.~A.}\ \bibnamefont
  {Carlson}},\ }\href {\doibase 10.3354/meps09985} {\bibfield  {journal}
  {\bibinfo  {journal} {Marine Ecology Progress Series}\ }\textbf {\bibinfo
  {volume} {470}},\ \bibinfo {pages} {249} (\bibinfo {year} {2012})},\ \bibinfo
  {note} {00006}\BibitemShut {NoStop}%
\bibitem [{\citenamefont {McDonnell}\ and\ \citenamefont
  {Buesseler}(2010)}]{mcdonnell_variability_2010}%
  \BibitemOpen
  \bibfield  {author} {\bibinfo {author} {\bibfnamefont {A.~M.~P.}\
  \bibnamefont {McDonnell}}\ and\ \bibinfo {author} {\bibfnamefont {K.~O.}\
  \bibnamefont {Buesseler}},\ }\href {\doibase 10.4319/lo.2010.55.5.2085}
  {\bibfield  {journal} {\bibinfo  {journal} {Limnology and Oceanography}\
  }\textbf {\bibinfo {volume} {55}},\ \bibinfo {pages} {2085} (\bibinfo {year}
  {2010})}\BibitemShut {NoStop}%
\bibitem [{\citenamefont {Petrik}\ \emph {et~al.}(2013)\citenamefont {Petrik},
  \citenamefont {Jackson},\ and\ \citenamefont
  {Checkley~Jr.}}]{petrik_aggregates_2013}%
  \BibitemOpen
  \bibfield  {author} {\bibinfo {author} {\bibfnamefont {C.~M.}\ \bibnamefont
  {Petrik}}, \bibinfo {author} {\bibfnamefont {G.~A.}\ \bibnamefont {Jackson}},
  \ and\ \bibinfo {author} {\bibfnamefont {D.~M.}\ \bibnamefont
  {Checkley~Jr.}},\ }\href {\doibase 10.1016/j.dsr.2012.12.009} {\bibfield
  {journal} {\bibinfo  {journal} {Deep Sea Research Part I: Oceanographic
  Research Papers}\ }\textbf {\bibinfo {volume} {74}},\ \bibinfo {pages} {64}
  (\bibinfo {year} {2013})}\BibitemShut {NoStop}%
\bibitem [{\citenamefont {Gargett}(1989)}]{gargett_ocean_1989}%
  \BibitemOpen
  \bibfield  {author} {\bibinfo {author} {\bibfnamefont {A.~E.}\ \bibnamefont
  {Gargett}},\ }\href {\doibase 10.1146/annurev.fl.21.010189.002223} {\bibfield
   {journal} {\bibinfo  {journal} {Annual Review of Fluid Mechanics}\ }\textbf
  {\bibinfo {volume} {21}},\ \bibinfo {pages} {419} (\bibinfo {year}
  {1989})}\BibitemShut {NoStop}%
\bibitem [{\citenamefont {Murray}(1970)}]{murray_settling_1970}%
  \BibitemOpen
  \bibfield  {author} {\bibinfo {author} {\bibfnamefont {S.~P.}\ \bibnamefont
  {Murray}},\ }\href {\doibase 10.1029/JC075i009p01647} {\bibfield  {journal}
  {\bibinfo  {journal} {Journal of Geophysical Research}\ }\textbf {\bibinfo
  {volume} {75}},\ \bibinfo {pages} {1647} (\bibinfo {year}
  {1970})}\BibitemShut {NoStop}%
\bibitem [{\citenamefont {Tooby}\ \emph {et~al.}(1977)\citenamefont {Tooby},
  \citenamefont {Wick},\ and\ \citenamefont {Isaacs}}]{tooby_motion_1977}%
  \BibitemOpen
  \bibfield  {author} {\bibinfo {author} {\bibfnamefont {P.~F.}\ \bibnamefont
  {Tooby}}, \bibinfo {author} {\bibfnamefont {G.~L.}\ \bibnamefont {Wick}}, \
  and\ \bibinfo {author} {\bibfnamefont {J.~D.}\ \bibnamefont {Isaacs}},\
  }\href {\doibase 10.1029/JC082i015p02096} {\bibfield  {journal} {\bibinfo
  {journal} {Journal of Geophysical Research}\ }\textbf {\bibinfo {volume}
  {82}},\ \bibinfo {pages} {2096} (\bibinfo {year} {1977})}\BibitemShut
  {NoStop}%
\bibitem [{\citenamefont {Alldredge}\ and\ \citenamefont
  {Gotschalk}(1988)}]{alldredge_situ_1988}%
  \BibitemOpen
  \bibfield  {author} {\bibinfo {author} {\bibfnamefont {A.~L.}\ \bibnamefont
  {Alldredge}}\ and\ \bibinfo {author} {\bibfnamefont {C.}~\bibnamefont
  {Gotschalk}},\ }\href {\doibase 10.4319/lo.1988.33.3.0339} {\bibfield
  {journal} {\bibinfo  {journal} {Limnology and Oceanography}\ }\textbf
  {\bibinfo {volume} {33}},\ \bibinfo {pages} {339} (\bibinfo {year}
  {1988})}\BibitemShut {NoStop}%
\bibitem [{\citenamefont {Shanks}(2002)}]{shanks_abundance_2002}%
  \BibitemOpen
  \bibfield  {author} {\bibinfo {author} {\bibfnamefont {A.~L.}\ \bibnamefont
  {Shanks}},\ }\href {\doibase 10.1016/S0278-4343(02)00015-8} {\bibfield
  {journal} {\bibinfo  {journal} {Continental Shelf Research}\ }\textbf
  {\bibinfo {volume} {22}},\ \bibinfo {pages} {2045} (\bibinfo {year}
  {2002})}\BibitemShut {NoStop}%
\bibitem [{\citenamefont {Ruiz}\ \emph {et~al.}(2004)\citenamefont {Ruiz},
  \citenamefont {Macías},\ and\ \citenamefont
  {Peters}}]{ruiz_turbulence_2004}%
  \BibitemOpen
  \bibfield  {author} {\bibinfo {author} {\bibfnamefont {J.}~\bibnamefont
  {Ruiz}}, \bibinfo {author} {\bibfnamefont {D.}~\bibnamefont {Macías}}, \
  and\ \bibinfo {author} {\bibfnamefont {F.}~\bibnamefont {Peters}},\ }\href
  {\doibase 10.1073/pnas.0401539101} {\bibfield  {journal} {\bibinfo  {journal}
  {Proceedings of the National Academy of Sciences of the United States of
  America}\ }\textbf {\bibinfo {volume} {101}},\ \bibinfo {pages} {17720}
  (\bibinfo {year} {2004})}\BibitemShut {NoStop}%
\bibitem [{Note1()}]{Note1}%
  \BibitemOpen
  \bibinfo {note} {The effective density is not easy to determine directly, but
  is often inferred from measured settling velocities based on the assumption
  of the validity of the Stokes law or of a modification thereof.}\BibitemShut
  {Stop}%
\bibitem [{\citenamefont {Kranenburg}(1994)}]{kranenburg_fractal_1994}%
  \BibitemOpen
  \bibfield  {author} {\bibinfo {author} {\bibfnamefont {C.}~\bibnamefont
  {Kranenburg}},\ }\href {\doibase 10.1016/S0272-7714(06)80002-8} {\bibfield
  {journal} {\bibinfo  {journal} {Estuarine, Coastal and Shelf Science}\
  }\textbf {\bibinfo {volume} {39}},\ \bibinfo {pages} {451} (\bibinfo {year}
  {1994})}\BibitemShut {NoStop}%
\bibitem [{\citenamefont {Winterwerp}(1998)}]{winterwerp_simple_1998}%
  \BibitemOpen
  \bibfield  {author} {\bibinfo {author} {\bibfnamefont {J.~C.}\ \bibnamefont
  {Winterwerp}},\ }\href {\doibase 10.1080/00221689809498621} {\bibfield
  {journal} {\bibinfo  {journal} {Journal of Hydraulic Research}\ }\textbf
  {\bibinfo {volume} {36}},\ \bibinfo {pages} {309} (\bibinfo {year}
  {1998})}\BibitemShut {NoStop}%
\bibitem [{\citenamefont {Logan}\ and\ \citenamefont
  {Wilkinson}(1990)}]{logan_fractal_1990}%
  \BibitemOpen
  \bibfield  {author} {\bibinfo {author} {\bibfnamefont {B.~E.}\ \bibnamefont
  {Logan}}\ and\ \bibinfo {author} {\bibfnamefont {D.~B.}\ \bibnamefont
  {Wilkinson}},\ }\href {\doibase 10.4319/lo.1990.35.1.0130} {\bibfield
  {journal} {\bibinfo  {journal} {Limnology and Oceanography}\ }\textbf
  {\bibinfo {volume} {35}},\ \bibinfo {pages} {130} (\bibinfo {year}
  {1990})}\BibitemShut {NoStop}%
\bibitem [{\citenamefont {Khelifa}\ and\ \citenamefont
  {Hill}(2006)}]{khelifa_models_2006}%
  \BibitemOpen
  \bibfield  {author} {\bibinfo {author} {\bibfnamefont {A.}~\bibnamefont
  {Khelifa}}\ and\ \bibinfo {author} {\bibfnamefont {P.~S.}\ \bibnamefont
  {Hill}},\ }\href {\doibase 10.1080/00221686.2006.9521690} {\bibfield
  {journal} {\bibinfo  {journal} {Journal of Hydraulic Research}\ }\textbf
  {\bibinfo {volume} {44}},\ \bibinfo {pages} {390} (\bibinfo {year}
  {2006})}\BibitemShut {NoStop}%
\bibitem [{\citenamefont {Jackson}\ \emph {et~al.}(2005)\citenamefont
  {Jackson}, \citenamefont {Waite},\ and\ \citenamefont
  {Boyd}}]{jackson_role_2005}%
  \BibitemOpen
  \bibfield  {author} {\bibinfo {author} {\bibfnamefont {G.~A.}\ \bibnamefont
  {Jackson}}, \bibinfo {author} {\bibfnamefont {A.~M.}\ \bibnamefont {Waite}},
  \ and\ \bibinfo {author} {\bibfnamefont {P.~W.}\ \bibnamefont {Boyd}},\
  }\href {\doibase 10.1029/2005GL023180} {\bibfield  {journal} {\bibinfo
  {journal} {Geophysical Research Letters}\ }\textbf {\bibinfo {volume} {32}},\
  \bibinfo {pages} {L13607} (\bibinfo {year} {2005})}\BibitemShut {NoStop}%
\bibitem [{\citenamefont {Zahnow}\ \emph {et~al.}(2011)\citenamefont {Zahnow},
  \citenamefont {Maerz},\ and\ \citenamefont
  {Feudel}}]{zahnow_particle-based_2011}%
  \BibitemOpen
  \bibfield  {author} {\bibinfo {author} {\bibfnamefont {J.~C.}\ \bibnamefont
  {Zahnow}}, \bibinfo {author} {\bibfnamefont {J.}~\bibnamefont {Maerz}}, \
  and\ \bibinfo {author} {\bibfnamefont {U.}~\bibnamefont {Feudel}},\
  }\href@noop {} {\bibfield  {journal} {\bibinfo  {journal} {Physica D:
  Nonlinear Phenomena}\ }\textbf {\bibinfo {volume} {240}},\ \bibinfo {pages}
  {882} (\bibinfo {year} {2011})},\ \bibinfo {note} {00004}\BibitemShut
  {NoStop}%
\bibitem [{\citenamefont {McCave}(1984)}]{mccave_size_1984}%
  \BibitemOpen
  \bibfield  {author} {\bibinfo {author} {\bibfnamefont {I.~N.}\ \bibnamefont
  {McCave}},\ }\href {\doibase 10.1016/0198-0149(84)90088-8} {\bibfield
  {journal} {\bibinfo  {journal} {Deep Sea Research Part A. Oceanographic
  Research Papers}\ }\textbf {\bibinfo {volume} {31}},\ \bibinfo {pages} {329}
  (\bibinfo {year} {1984})}\BibitemShut {NoStop}%
\bibitem [{\citenamefont {Tambo}\ and\ \citenamefont
  {Watanabe}(1979)}]{tambo_physical_1979}%
  \BibitemOpen
  \bibfield  {author} {\bibinfo {author} {\bibfnamefont {N.}~\bibnamefont
  {Tambo}}\ and\ \bibinfo {author} {\bibfnamefont {Y.}~\bibnamefont
  {Watanabe}},\ }\href {\doibase 10.1016/0043-1354(79)90033-2} {\bibfield
  {journal} {\bibinfo  {journal} {Water Research}\ }\textbf {\bibinfo {volume}
  {13}},\ \bibinfo {pages} {409} (\bibinfo {year} {1979})}\BibitemShut
  {NoStop}%
\bibitem [{\citenamefont {Simon}\ \emph {et~al.}(2002)\citenamefont {Simon},
  \citenamefont {Grossart}, \citenamefont {Schweitzer},\ and\ \citenamefont
  {Ploug}}]{simon_microbial_2002}%
  \BibitemOpen
  \bibfield  {author} {\bibinfo {author} {\bibfnamefont {M.}~\bibnamefont
  {Simon}}, \bibinfo {author} {\bibfnamefont {H.~P.}\ \bibnamefont {Grossart}},
  \bibinfo {author} {\bibfnamefont {B.}~\bibnamefont {Schweitzer}}, \ and\
  \bibinfo {author} {\bibfnamefont {H.}~\bibnamefont {Ploug}},\ }\href@noop {}
  {\bibfield  {journal} {\bibinfo  {journal} {Aquatic microbial ecology}\
  }\textbf {\bibinfo {volume} {28}},\ \bibinfo {pages} {175} (\bibinfo {year}
  {2002})}\BibitemShut {NoStop}%
\bibitem [{\citenamefont {Soulsby}\ \emph {et~al.}(2013)\citenamefont
  {Soulsby}, \citenamefont {Manning}, \citenamefont {Spearman},\ and\
  \citenamefont {Whitehouse}}]{soulsby_settling_2013}%
  \BibitemOpen
  \bibfield  {author} {\bibinfo {author} {\bibfnamefont {R.~L.}\ \bibnamefont
  {Soulsby}}, \bibinfo {author} {\bibfnamefont {A.~J.}\ \bibnamefont
  {Manning}}, \bibinfo {author} {\bibfnamefont {J.}~\bibnamefont {Spearman}}, \
  and\ \bibinfo {author} {\bibfnamefont {R.~J.~S.}\ \bibnamefont
  {Whitehouse}},\ }\href {\doibase 10.1016/j.margeo.2013.04.006} {\bibfield
  {journal} {\bibinfo  {journal} {Marine Geology}\ }\textbf {\bibinfo {volume}
  {339}},\ \bibinfo {pages} {1} (\bibinfo {year} {2013})}\BibitemShut {NoStop}%
\bibitem [{\citenamefont {Winterwerp}(2011)}]{winterwerp_physical_2011}%
  \BibitemOpen
  \bibfield  {author} {\bibinfo {author} {\bibfnamefont {J.~C.}\ \bibnamefont
  {Winterwerp}},\ }in\ \href
  {http://www.sciencedirect.com/science/article/pii/B978012374711200214X}
  {\emph {\bibinfo {booktitle} {Treatise on {Estuarine} and {Coastal}
  {Science}}}},\ \bibinfo {editor} {edited by\ \bibinfo {editor} {\bibfnamefont
  {E.}~\bibnamefont {Wolanski}}\ and\ \bibinfo {editor} {\bibfnamefont
  {D.}~\bibnamefont {McLusky}}}\ (\bibinfo  {publisher} {Academic Press},\
  \bibinfo {address} {Waltham},\ \bibinfo {year} {2011})\ pp.\ \bibinfo {pages}
  {311--360}\BibitemShut {NoStop}%
\bibitem [{\citenamefont {Oakey}\ and\ \citenamefont
  {Elliott}(1982)}]{oakey_dissipation_1982}%
  \BibitemOpen
  \bibfield  {author} {\bibinfo {author} {\bibfnamefont {N.~S.}\ \bibnamefont
  {Oakey}}\ and\ \bibinfo {author} {\bibfnamefont {J.~A.}\ \bibnamefont
  {Elliott}},\ }\href@noop {} {\bibfield  {journal} {\bibinfo  {journal}
  {Journal of Physical Oceanography}\ }\textbf {\bibinfo {volume} {12}},\
  \bibinfo {pages} {171} (\bibinfo {year} {1982})}\BibitemShut {NoStop}%
\bibitem [{\citenamefont {Kirboe}\ and\ \citenamefont
  {Saiz}(1995)}]{kirboe_planktivorous_1995}%
  \BibitemOpen
  \bibfield  {author} {\bibinfo {author} {\bibfnamefont {T.}~\bibnamefont
  {Kirboe}}\ and\ \bibinfo {author} {\bibfnamefont {E.}~\bibnamefont {Saiz}},\
  }\href {\doibase 10.3354/meps122135} {\bibfield  {journal} {\bibinfo
  {journal} {Marine Ecology Progress Series}\ }\textbf {\bibinfo {volume}
  {122}},\ \bibinfo {pages} {135} (\bibinfo {year} {1995})}\BibitemShut
  {NoStop}%
\bibitem [{\citenamefont {Bartholomä}\ \emph {et~al.}(2009)\citenamefont
  {Bartholomä}, \citenamefont {Kubicki}, \citenamefont {Badewien},\ and\
  \citenamefont {Flemming}}]{bartholoma_suspended_2009}%
  \BibitemOpen
  \bibfield  {author} {\bibinfo {author} {\bibfnamefont {A.}~\bibnamefont
  {Bartholomä}}, \bibinfo {author} {\bibfnamefont {A.}~\bibnamefont
  {Kubicki}}, \bibinfo {author} {\bibfnamefont {T.~H.}\ \bibnamefont
  {Badewien}}, \ and\ \bibinfo {author} {\bibfnamefont {B.~W.}\ \bibnamefont
  {Flemming}},\ }\href@noop {} {\bibfield  {journal} {\bibinfo  {journal}
  {Ocean Dynamics}\ }\textbf {\bibinfo {volume} {59}},\ \bibinfo {pages} {213}
  (\bibinfo {year} {2009})},\ \bibinfo {note} {00029}\BibitemShut {NoStop}%
\bibitem [{Note2()}]{Note2}%
  \BibitemOpen
  \bibinfo {note} {For small excess densities $\Delta \rho $, characteristic to
  our cases, $\beta =1-\protect \frac {2}{3}\protect \frac {\Delta \rho }{\rho
  _f}$, as follows from Eq.(~\ref {beta}) for small $\Delta \rho /\rho
  _f$.}\BibitemShut {Stop}%
\bibitem [{\citenamefont {Pope}(2000)}]{pope_turbulent_2000}%
  \BibitemOpen
  \bibfield  {author} {\bibinfo {author} {\bibfnamefont {S.~B.}\ \bibnamefont
  {Pope}},\ }\href@noop {} {\emph {\bibinfo {title} {Turbulent {Flows}}}},\
  \bibinfo {edition} {1st}\ ed.\ (\bibinfo  {publisher} {Cambridge University
  Press},\ \bibinfo {year} {2000})\BibitemShut {NoStop}%
\bibitem [{\citenamefont {Canuto}\ \emph {et~al.}(1987)\citenamefont {Canuto},
  \citenamefont {Hussaini}, \citenamefont {Quarteroni},\ and\ \citenamefont
  {Zang}}]{canuto_spectral_1987}%
  \BibitemOpen
  \bibfield  {author} {\bibinfo {author} {\bibfnamefont {C.}~\bibnamefont
  {Canuto}}, \bibinfo {author} {\bibfnamefont {M.~Y.}\ \bibnamefont
  {Hussaini}}, \bibinfo {author} {\bibfnamefont {A.}~\bibnamefont
  {Quarteroni}}, \ and\ \bibinfo {author} {\bibfnamefont {T.~A.}\ \bibnamefont
  {Zang}},\ }\href {http://link.springer.com/10.1007/978-3-642-84108-8} {\emph
  {\bibinfo {title} {Spectral {Methods} in {Fluid} {Dynamics}}}}\ (\bibinfo
  {publisher} {Springer Berlin Heidelberg},\ \bibinfo {address} {Berlin,
  Heidelberg},\ \bibinfo {year} {1987})\BibitemShut {NoStop}%
\bibitem [{\citenamefont {Hou}\ and\ \citenamefont
  {Li}(2007)}]{hou_computing_2007}%
  \BibitemOpen
  \bibfield  {author} {\bibinfo {author} {\bibfnamefont {T.~Y.}\ \bibnamefont
  {Hou}}\ and\ \bibinfo {author} {\bibfnamefont {R.}~\bibnamefont {Li}},\
  }\href {\doibase 10.1016/j.jcp.2007.04.014} {\bibfield  {journal} {\bibinfo
  {journal} {Journal of Computational Physics}\ }\textbf {\bibinfo {volume}
  {226}},\ \bibinfo {pages} {379} (\bibinfo {year} {2007})}\BibitemShut
  {NoStop}%
\bibitem [{\citenamefont {Shu}\ and\ \citenamefont
  {Osher}(1988)}]{shu_efficient_1988}%
  \BibitemOpen
  \bibfield  {author} {\bibinfo {author} {\bibfnamefont {C.-W.}\ \bibnamefont
  {Shu}}\ and\ \bibinfo {author} {\bibfnamefont {S.}~\bibnamefont {Osher}},\
  }\href {\doibase 10.1016/0021-9991(88)90177-5} {\bibfield  {journal}
  {\bibinfo  {journal} {Journal of Computational Physics}\ }\textbf {\bibinfo
  {volume} {77}},\ \bibinfo {pages} {439} (\bibinfo {year} {1988})}\BibitemShut
  {NoStop}%
\bibitem [{Note3()}]{Note3}%
  \BibitemOpen
  \bibinfo {note} {Given the same initial condition on the same computer (i.e.
  floating-point architecture) with a fixed time step, would not lead to a
  separation of two runs even in chaos with any chosen equation of motion. Note
  that the deviations generated by our three different dynamics can be
  additionally amplified by the chaotic nature of the particle dynamics. Here,
  were are unable to separate out this amplification.}\BibitemShut {Stop}%
\bibitem [{Note4()}]{Note4}%
  \BibitemOpen
  \bibinfo {note} {The limit of $N_p \rightarrow \infty $ can only be
  considered for mathematical convenience. In the case of finite size particles
  it is important to remain in the dilute limit, in order to avoid hydrodynamic
  interactions.}\BibitemShut {Stop}%
\bibitem [{\citenamefont {Batchelor}(1950)}]{batchelor_application_1950}%
  \BibitemOpen
  \bibfield  {author} {\bibinfo {author} {\bibfnamefont {G.~K.}\ \bibnamefont
  {Batchelor}},\ }\href {\doibase 10.1002/qj.49707632804} {\bibfield  {journal}
  {\bibinfo  {journal} {Quarterly Journal of the Royal Meteorological Society}\
  }\textbf {\bibinfo {volume} {76}},\ \bibinfo {pages} {133} (\bibinfo {year}
  {1950})}\BibitemShut {NoStop}%
\bibitem [{\citenamefont {Wang}\ and\ \citenamefont
  {Maxey}(1993)}]{wang_settling_1993}%
  \BibitemOpen
  \bibfield  {author} {\bibinfo {author} {\bibfnamefont {L.-P.}\ \bibnamefont
  {Wang}}\ and\ \bibinfo {author} {\bibfnamefont {M.~R.}\ \bibnamefont
  {Maxey}},\ }\href {\doibase 10.1017/S0022112093002708} {\bibfield  {journal}
  {\bibinfo  {journal} {Journal of Fluid Mechanics}\ }\textbf {\bibinfo
  {volume} {256}},\ \bibinfo {pages} {27} (\bibinfo {year} {1993})}\BibitemShut
  {NoStop}%
\bibitem [{\citenamefont {Maxey}(1987)}]{maxey_gravitational_1987}%
  \BibitemOpen
  \bibfield  {author} {\bibinfo {author} {\bibfnamefont {M.~R.}\ \bibnamefont
  {Maxey}},\ }\href {\doibase 10.1017/S0022112087000193} {\bibfield  {journal}
  {\bibinfo  {journal} {Journal of Fluid Mechanics}\ }\textbf {\bibinfo
  {volume} {174}},\ \bibinfo {pages} {441} (\bibinfo {year}
  {1987})}\BibitemShut {NoStop}%
\bibitem [{Note5()}]{Note5}%
  \BibitemOpen
  \bibinfo {note} {The excess density is important, of course, for $W_{\protect
  \text {settling}}$, see also (\ref {W}), but not for the ratio investigated
  here.}\BibitemShut {Stop}%
\end{thebibliography}%

\end{document}